\documentclass[onecolumn, journal]{IEEEtran}
\usepackage{graphicx}
\usepackage{tikz}
\usepackage{pgfplots}
\usetikzlibrary{calc}
\usetikzlibrary{fit}
\pgfplotsset{compat=newest}

\usetikzlibrary{shapes,arrows,fit,positioning}
\usetikzlibrary{plotmarks}
\usetikzlibrary{decorations.pathreplacing}
\usetikzlibrary{patterns}

\tikzstyle{block} = [draw, fill=white, rectangle,minimum height=3em, minimum width=5.2em]	% Block diagrams
\tikzstyle{block_rot} = [draw, fill=white, rectangle,minimum height=5.2em, minimum width=3em]
\tikzstyle{sum} = [draw, fill=white, circle, node distance=1em,path picture={\draw[black](path picture bounding box.south) -- (path picture bounding box.north) (path picture bounding box.west) -- (path picture bounding box.east);}]
\tikzstyle{coord} = [coordinate]

\tikzstyle{state}=[shape=circle,draw=blueTUD50,fill=blueTUD10]	% Trellis states
\tikzstyle{lightedge}=[<-,dotted]
\tikzstyle{mainstate}=[state,thick]
\tikzstyle{mainedge}=[<-,thick]

\tikzstyle{symbol}=[shape=circle,draw=blueTUD50,fill=blueTUD10,minimum width=1em,scale=0.6]	% Symbols
\tikzstyle{sample1}=[shape=circle,draw,scale=0.3]	% Samples
\tikzstyle{sample2}=[shape=circle,draw,densely dashed,scale=0.3]	% Samples
\tikzstyle{interleave}=[shape=circle,draw,fill,minimum width=1em,scale=0.6]	% Interleaver

\tikzstyle{register} = [draw, fill=white, rectangle,minimum height=3em, minimum width=3em]	% Block diagrams
\tikzstyle{mod2} = [draw, fill=white, circle, label={mod 2}, node distance=1em,path picture={\draw[black](path picture bounding box.south) -- (path picture bounding box.north) (path picture bounding box.west) -- (path picture bounding box.east);}]

\tikzset{>=latex}

\pgfplotsset{
	every axis/.append style = {
		grid = major,
	},
	every axis legend/.append style	={
		font = \small,
		legend cell align = left
	}
}
	\definecolor{lbrown}{rgb}{0.86, 0.7, 0.52}		
	
	\definecolor{dblue}{rgb}{0.08, 0.38, 0.74}
	\definecolor{lblue}{rgb}{0.7 , 0.7 , 1}
	\definecolor{ldblue}{rgb}{0.39, 0.58, 0.93}
	
	\definecolor{damber}{rgb}{0.72, 0.53, 0.04}
	\definecolor{lamber}{rgb}{1.0, 0.9, 0.6}
	\definecolor{amber}{rgb}{1.0, 0.75, 0.0}
	\definecolor{ldamber}{rgb}{0.98, 0.85, 0.37}
	
	\definecolor{lgray}{rgb}{0.8, 0.85, 0.81}
	\definecolor{ashgrey}{rgb}{0.7, 0.75, 0.71}
	
	\definecolor{dred}{rgb}{0.8, 0.0, 0.0}
	\definecolor{lred}{rgb}{1, 0.7 , 0.7 }
	\definecolor{ddred}{rgb}{0.5, 0.0, 0.13}
	\definecolor{ldred}{rgb}{0.89, 0.44, 0.48}
	\definecolor{lgreen}{rgb}{0.6, 0.75, 0.6}
	\definecolor{ldgreen}{rgb}{0.4, 0.7, 0.4}
	\definecolor{dgreen}{rgb}{0.0, 0.5, 0.0}
	\definecolor{ddgreen}{rgb}{0.12, 0.3, 0.17}
	\definecolor{applegreen}{rgb}{0.55, 0.71, 0.0}
	\definecolor{brightorange}{rgb}{1.0, 0.56, 0.0}

\usepackage{xfrac}
\usepackage{amssymb}
\usepackage{trfsigns}
\usepackage{bm}
\usepackage{bbm}
\usepackage[cmex10]{amsmath}

\DeclareMathOperator\erfc{erfc}

\DeclareMathOperator\Var{Var}
\DeclareMathOperator\Ci{Ci}
\DeclareMathOperator\Si{Si}

\DeclareMathOperator\arcosh{arcosh}
\DeclareMathOperator\E{\mathbb{E}}
\newcommand{\rv}[1]{\mathsf{#1}}
\newcommand{\V}[1]{\boldsymbol{#1}}
\newcommand{\Vrv}[1]{\boldsymbol{\mathsf{#1}}}
\newcommand\setItemnumber[1]{\setcounter{enumi}{\numexpr#1-1\relax}}

\usepackage{cite}

\usepackage{enumitem}

\begin{document}
\title{On the Achievable Rate of Bandlimited Continuous-Time AWGN Channels with 1-Bit Output Quantization\thanks{This work was supported in part by the German Research Foundation (DFG) in the Collaborative Research Center SFB912, "Highly Adaptive Energy-Efficient Computing", HAEC. Parts of this work have been presented at the IEEE International Symposium on Information Theory (ISIT), Aachen, Germany, June, 2017 \cite{Bender2017_lowerBound} and have been submitted to the International Zurich Seminar on Information and Communication (IZS), Zurich, Switzerland, February, 2018 \cite{Bender2017_upperBound}. This work has been submitted to the IEEE for possible publication.  Copyright may be transferred without notice, after which this version may no longer be accessible.}}

\author{Sandra~Bender, 
        Meik~D\"orpinghaus, 
        and~Gerhard~Fettweis % <-this % stops a space
\thanks{S. Bender, M. D\"orpinghaus and G. Fettweis are with the Vodafone Chair Mobile Communications Systems, Faculty of Electrical and Computer Engineering, Technische Universit\"at Dresden, 01062 Dresden, Germany. E-mail: $\{$sandra.bender, meik.doerpinghaus, gerhard.fettweis$\}$@tu-dresden.de}% <-this % stops a space
}

\maketitle

\begin{abstract}
	We consider a continuous-time bandlimited additive white Gaussian noise channel with 1-bit output quantization. On such a channel the information is carried by the temporal distances of the zero-crossings of the transmit signal. The set of input signals is constrained by the bandwidth of the channel and an average power constraint. We derive a lower bound on the capacity by lower-bounding the mutual information rate for a given set of waveforms with exponentially distributed zero-crossing distances, where we focus on the behavior in the mid to high signal-to-noise ratio regime.
	We find that in case the input randomness scales appropriately with the available bandwidth, the mutual information rate grows linearly with the channel bandwidth for constant signal-to-noise ratios. Furthermore, for a given bandwidth the lower bound saturates with the signal-to-noise ratio growing to infinity. The ratio between the lower bound on the mutual information rate and the capacity of the additive white Gaussian noise channel without quantization is a constant independent of the channel bandwidth for an appropriately chosen randomness of the channel input and a given signal-to-noise ratio.
	We complement those findings with an upper bound on the mutual information rate for the specific signaling scheme. We show that both bounds are close in the mid to high SNR domain.
\end{abstract}

\begin{IEEEkeywords}
\vspace{-2mm}
channel capacity, one-bit quantization, timing channel, continuous-time channel
\end{IEEEkeywords}

\IEEEpeerreviewmaketitle

\section{Introduction}
In digital communications, we typically assume that the analog-to-digital converter (ADC) at the receiver provides a sufficiently fine grained quantization of the magnitude of the received signal.
However, for very high data rate short link communication the power consumption of the ADC becomes a major factor, also in comparison to the transmit power. This is due to the required high quantization resolution at a very high sampling rate and the fact that the consumed energy per conversion step increases with the sampling rate \cite{Murmann_Survey_16}.
One idea to circumvent this problem is the use of 1-bit quantization and oversampling of the received signal w.r.t. the Nyquist rate. One-bit quantization is fairly simple to realize as no highly linear analog signal processing is required. Quantization resolution of the signal magnitude is then traded-off by resolution in time domain. Optimal communication over the resulting channel including the ADC requires a modulation and signaling scheme adapted to this specific channel as the information is no longer carried in the signal magnitude but in the zero-crossing time instants of the transmitted signal. The question then is, how much the channel capacity is degraded compared to an additive white Gaussian noise (AWGN) channel quantized with high resolution and sampled at Nyquist rate.\looseness-1

For the noise free case it has been shown already in the early works by Gilbert~\cite{Gilbert1993} and Shamai~\cite{Shamai1994} that oversampling of a bandlimited channel can increase the information rate w.r.t. Nyquist sampling. The latter lower-bounded the capacity by $\log_2 (n +1)$\,[bits/Nyquist interval] where $n$ is the oversampling factor w.r.t. Nyquist sampling. 
Regarding the low signal-to-noise ratio (SNR) domain, Koch and Lapidoth have shown in \cite{KochLapidoth10} that oversampling increases the capacity per unit-cost of bandlimited Gaussian channels with 1-bit output quantization. In \cite{zhang2012} it has been shown that oversampling increases the achievable rate based on the study of the generalized mutual information. 
Moreover, in \cite{landau2014information} simulative approaches on bounding the achievable rate in a discrete-time scenario are studied. In \cite{landau2014reconstructable,Landau2015,Bender2016}, the achievable rate is evaluated via simulation for different signaling strategies.

However, an analytical evaluation of the channel capacity of the 1-bit quantized oversampled AWGN channel in the mid to high SNR domain is still open. This capacity depends on the oversampling factor, as due to the 1-bit quantization Nyquist-sampling, like any other sampling rate, does not provide a sufficient statistic. As a limiting case, we study the capacity of the underlying continuous-time 1-bit quantized channel, which corresponds to the case where the oversampling factor becomes infinitely large. Without time quantization, there is, as for the capacity of the AWGN channel as given by Shannon \cite{Shannon1948}, no quantization in the information carrying dimension. However, the capacity of the AWGN channel without output quantization is an upper bound on the capacity of the continuous-time 1-bit quantized channel. 
With our approach, we aim for a better understanding of the difference between using the magnitude domain versus the time domain for signaling. As the continuous-time additive noise channel with 1-bit output quantization
carries the information in the zero-crossings of the transmit signal, this channel corresponds to some extent to a timing channel as, e.g., studied in \cite{AnantharamVerdu1996}.\looseness-1

Given the outlined application scenario of short range multigigabit/s-communication, we focus on the mid to high SNR domain. Firstly, we derive a lower bound on the mutual information rate of the bandlimited continuous-time additive Gaussian noise channel with 1-bit output quantization. We show that the mutual information rate increases with the bandwidth for an appropriately chosen input distribution but saturates over the SNR. Moreover, we observe that the ratio between our lower bound and the AWGN capacity is a constant independent of the bandwidth for a given SNR and the appropriately chosen input distribution mentioned before.
Secondly, we derive the corresponding upper bound on the mutual information rate for the specific signaling scheme in order to quantify the deviation of the lower bound from the actual rate, which results from the applied bounding steps.
The derivations are based on certain approximations and simplifications, which will be clearly stated and are suitable in the mid to high SNR domain. We observe that both bounds are close in the mid to high SNR regime, where for a given input distribution the gaps size decreases with increasing bandwidth.

The rest of the paper is organized as follows. In Section~\ref{sec:Intro}, the system model is introduced and the different types of error events as well as the set of assumptions applied are discussed. Based on those, an upper and a lower bound on the mutual information rate are given in Section~\ref{sec:error_events} and analyzed in detail in Sections~\ref{sec:genie-aided-rx} and \ref{sec:Aux_Proc}.
	Subsequently, in Section~\ref{sec:LP-distortion} the effects of the	distortion introduced by the transmit and receive filters are discussed. In Section~\ref{sec:Conclusion} we give the final form of the upper and the lower bound on the mutual information rate and discuss their behavior depending on various channel parameters. Section~\ref{sec:conclusion} provides the conclusion of our findings.

We apply the following notations: vectors are set bold, random variables sans serif. Thus, $\V{\rv{X}}^{(K)}$ is a random vector of length $K$. Omitting the superscript denotes the corresponding random process $\V{\rv{X}}$ for $K \rightarrow \infty$.  Upper and lower bounds are denoted by an upper bar and an underline, respectively. For information measures, $(\cdot)'$ denotes the corresponding rate. Hence, $\underline{I}'(\Vrv{X};\Vrv{Y})$ is a lower bound on the mutual information rate between $\Vrv{X}$ and $\Vrv{Y}$. Furthermore, $(a)^{+}$ is the maximum of $a$ and zero.
	
% % % % % % % % % % % % % % % % % % % % % % % % % % % % % % % % % % % % % % % % % % % % % % % % % % % % % % % % % % % % % % % % % % % % % % % % % % % % % % % % % % % % % % % % % % % % % % % % % % % % % % % % % % % % %
% --------------------------------------------------------------------------------------------------------------------------------------------------------------------------------------------------------------------- %
% % % % % % % % % % % % % % % % % % % % % % % % % % % % % % % % % % % % % % % % % % % % % % % % % % % % % % % % % % % % % % % % % % % % % % % % % % % % % % % % % % % % % % % % % % % % % % % % % % % % % % % % % % % % %
\section{System Model}
\label{sec:Intro}
	\begin{figure}[t]
		\centering
		\includegraphics{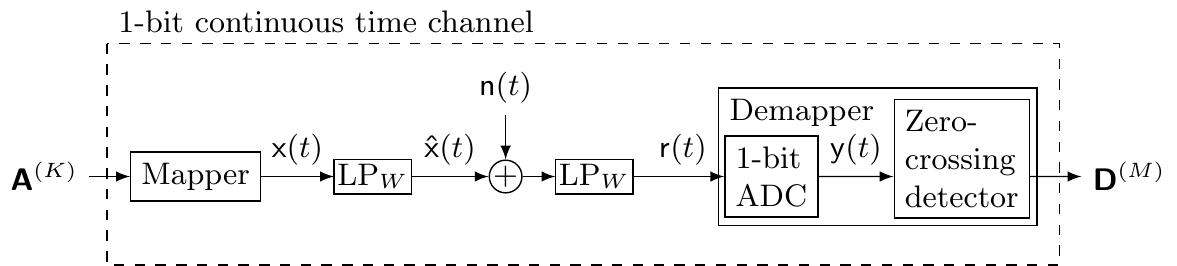}
		\caption{System model}
		\label{fig:SysMod1}
	\end{figure}
	
	We consider the system model depicted in Fig. \ref{fig:SysMod1}. A receiver relying on 1-bit quantization can only distinguish between the level of the input signal being smaller or larger than zero. Hence, all information that can be conveyed through such a channel is encoded in the time instants of the zero-crossings\footnote{Note that one additional bit is carried by the sign of the signal. However, its effect on the mutual information between channel input and output can be neglected when studying the capacity as it converges to zero for infinite blocklength.}. 
	In order to model this, we consider as channel input
	and output the vectors $\boldsymbol{\rv{A}}^{(K)} = [\rv{A}_1,...,\rv{A}_K]^T$ and $\boldsymbol{\rv{D}}^{(M)}=[\rv{D}_1,...,\rv{D}_M]^T$, which contain the temporal distances $\rv{A}_k$ and $\rv{D}_m$ of two consecutive zero-crossings~(ZC) of $\rv{x}(t)$ and the received signal $\rv{r}(t)$, respectively. Here $K$ is not necessarily equal to $M$ as noise can add
	or remove zero-crossings. 
	For the analysis in this work, it is assumed that the time instants of the zero crossings can be resolved with infinite precision, which makes $\rv{A}_k$ and $\rv{D}_m$ continuous random variables. The mapper converts the random vector $\boldsymbol{\rv{A}}^{(K)}$ into the  continuous-time transmit signal $\rv{x}(t)$, which is then lowpass-filtered with one-sided bandwidth $W$ and transmitted over an AWGN channel. At the receiver, lowpass-filtering with one-sided bandwidth $W$ ensures bandlimitation of the noise and the demapper realizes the conversion between the noisy received signal $\rv{r}(t)$ and the sequence $\boldsymbol{\rv{D}}^{(M)}$ of zero-crossing distances. 
% --------------------------------------------------------------------------------------------------------------------------------------------------------------------------------------------------------------------- %	
	\subsection{Signal Structure and Input Distribution}
	\label{subsec:System_Input}
\begin{figure}[t]
		\centering
		\includegraphics{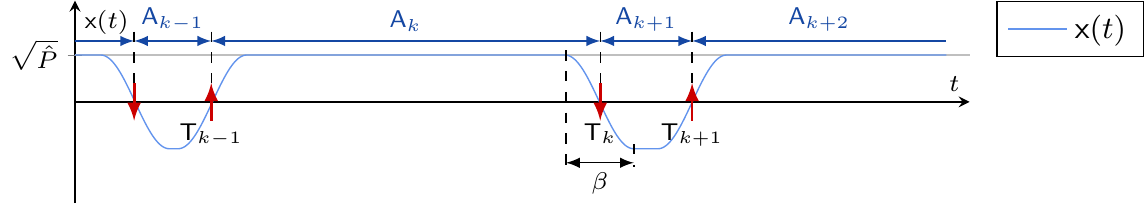}
		\caption{Mapping from input sequence $\boldsymbol{\rv{A}}^{(K)}$ to transmit signal $\rv{x}(t)$}
		\label{fig:input_mapping}
	\end{figure}

	Fig.~\ref{fig:input_mapping} illustrates the mapping of the input sequence $\boldsymbol{\rv{A}}^{(K)}$ to $\rv{x}(t)$, which alternates between two levels $\pm \sqrt{\hat{P}}$, where $\hat{P}$ is the peak power of the input signal. 
	The $k$th zero-crossing corresponding to $\rv{A}_k$ occurs at time 
	\begin{IEEEeqnarray}{rCl}
		\rv{T}_k = \sum\limits_{i=1}^{k} \rv{A}_i + t_0. \label{eq:Ak-Tk_relation}
	\end{IEEEeqnarray}
	Without loss of generality, we assume $t_0=0$. 
	The input symbols $\rv{A}_k$ correspond to the temporal distances between the $k$th and the $(k-1)$th zero-crossing of $\rv{x}(t)$. We consider i.i.d. exponentially distributed $\rv{A}_k$ with
	\begin{IEEEeqnarray}{rCl}
	\rv{A}_k \sim \lambda e^{-\lambda(a-\beta)} \mathbbm{1}_{\left[\beta,\infty\right)}(a) \label{eq:p_A(a)}
	\end{IEEEeqnarray}
	since the exponential distribution maximizes the entropy for positive continuous random variables with given mean.  Here, $\mathbbm{1}_{[u,v]}(x)$ is the indicator function, being one in the interval $[u,v]$ and zero otherwise. This results in a mean symbol duration of
	\begin{IEEEeqnarray}{rCl}
		T_{\text{avg}} = \frac{1}{\lambda} + \beta \label{eq:T_avg}
	\end{IEEEeqnarray}
	and a variance of the input symbols of
	\begin{IEEEeqnarray}{rCL}
		\sigma_\rv{A}^2 = \sfrac{1}{\lambda^2}. \label{eq:input_variance}
	\end{IEEEeqnarray}
	In order to control the bandwidth of the channel input signal and match it to the channel, the transition from one level to the other is given by the waveform $f(t)$, yielding the transmit signal
	\begin{IEEEeqnarray}{rCl}
	\rv{x}(t) = \left(\sum_{k=1}^{K} \sqrt{\hat{P}} (-1)^k g(t-\rv{T}_k)\right) + \sqrt{\hat{P}} \label{eq:x(t)}
	\end{IEEEeqnarray}
	with the pulse shape
	\begin{IEEEeqnarray}{rCl}
	g(t) = \left(1+f\left(t-\frac{\beta}{2}\right)\right)\!\cdot\!\mathbbm{1}_{\left[0,\beta\right]}(t) + 2\!\cdot\!\mathbbm{1}_{\left[\beta,\infty\right)}(t) \label{eq:g(t)_allgemein}
	\end{IEEEeqnarray}
	where $f(t)$ is an odd function between $(-\sfrac{\beta}{2},-1)$ and $(\sfrac{\beta}{2},1)$ and zero otherwise, describing the transition of the signal. The transition time $\beta$ is chosen according to the available channel bandwidth $W$ with 
	\begin{IEEEeqnarray}{rCl}
		\beta=\frac{1}{2 W}. \label{eq:Relation_W_beta}
	\end{IEEEeqnarray}
	Implications of this choice will be discussed in Section~\ref{subsec:Lowpass_A} and Section~\ref{sec:Conclusion}.
	With $\beta$ being the minimal value of the $\rv{A}_{k}$, it is guaranteed that $\rv{x}(t)$ achieves $\sqrt{\hat{P}}$ between two transitions. This is not necessarily capacity-achieving but simplifies the derivation of a lower bound on the mutual information rate, as we will discuss in Section~\ref{subsec:System_Assumptions}.
	If not stated otherwise, results throughout the paper are given for a sine halfwave as transition, i.e.,\looseness-1
	\begin{align}
		f(t) &= \begin{cases}
		\sin\left(\pi \frac{t}{\beta}\right) &\text{for } |t| \le \sfrac{\beta}{2}\\
		0 &\text{otherwise}
		\end{cases}. \label{eq:f(t)_cosine}
	\end{align}
	In the limiting case of $\lambda \rightarrow \infty$, this leads to a one sided signal bandwidth of $W$. 
	However, $\rv{x}(t)$ is not strictly bandlimited as a small portion of its energy is outside of the interval $[-W, W ]$.
	Strict bandlimitation is ensured by the lowpass (LP) filters at transmitter and receiver, which are considered to be ideal LPs with one-sided bandwidth $W$ {and amplitude one}.
% --------------------------------------------------------------------------------------------------------------------------------------------------------------------------------------------------------------------- %
	\subsection{Channel Model}
	\label{subsec:System_Channel}
	The {LP-filtered} signal $\rv{\hat{x}}(t)$ is transmitted over a continuous-time AWGN channel. The received signal after {quantization and LP-filtering} is given by
	\begin{IEEEeqnarray}{rCL}
		\rv{y}(t)=Q(\rv{x}(t)+\rv{z}(t))
	\end{IEEEeqnarray}
	where $Q(\cdot)$ denotes a binary quantizer with threshold zero, i.e., $Q(x) = 1$ if $x\ge0$ and $Q(x) = -1$ if $x<0$. Moreover, $\rv{z}(t)$ is the overall additive distortion between $\rv{x}(t)$ and the received signal $\rv{r}(t)$
	\begin{IEEEeqnarray}{rCL}
		\rv{z}(t) = \rv{r}(t)-\rv{x}(t) = \rv{\hat{n}}(t) + \rv{\tilde{x}}(t).
	\end{IEEEeqnarray}
	Here $\rv{\hat{n}(t)}$ is the filtered version of the zero-mean additive white Gaussian noise $\rv{n}(t)$ with power spectral density (PSD) $\sfrac{N_0}{2}$, its PSD is given by
	\begin{IEEEeqnarray}{rCL}
		S_{\rv{\hat{n}}}(f) = \begin{cases}
			\sfrac{N_0}{2} & \text{for } |f| \le W\\
			0 &\text{otherwise}
		\end{cases}
	\end{IEEEeqnarray}
	and its variance is $ \sigma_{\rv{\hat{n}}}^2 = N_0 W$.
	Furthermore $\rv{\tilde{x}}(t) = \rv{\hat{x}}(t)-\rv{x}(t)$ is the LP-distortion introduced by transmit and receive-filtering, which we model as an additional noise source. Its variance is given by 
	\begin{IEEEeqnarray}{rCL}
		\sigma_{\rv{\tilde{x}}}^2 = \E\big[\left|x(t)-\hat{x}(t)\right|^2\big] = \frac{1}{\pi} \int_{2 \pi W}^{\infty} S_{\rv{x}}(\omega) d\omega
	\end{IEEEeqnarray}
	where $S_{\rv{x}}(\omega)$ is the PSD of $\rv{x}(t)$. For the variance of the overall distortion $\rv{z}(t)$, we thus obtain
	\begin{IEEEeqnarray}{rCL}
		\sigma_\rv{z}^2 = \sigma_\rv{\hat{n}}^2 + \sigma_\rv{\tilde{x}}^2. \label{eq:var_z(t)}
	\end{IEEEeqnarray}
	The signal-to-noise ratio after the transmit filter, i.e., w.r.t. $\rv{\hat{x}}(t)$, is
	\begin{IEEEeqnarray}{rCL}
		\rho^{\ast} = \frac{P-{\sigma}^2_{\rv{\tilde{x}}}}{N_0 W} \label{eq:trueSNR}
	\end{IEEEeqnarray}
	where $P$ is the average power of $\rv{x}(t)$.
	It is given by
	\begin{IEEEeqnarray}{rCL}
		P\!=\!\frac{\hat{P}}{T_{\text{avg}}} \left(\int_{0}^{\beta}\!\cos^2\!\left(\frac{\pi}{\beta}t\right)\!dt\!+\!\frac{1}{\lambda}\right)\!= \frac{\frac12 + 2 W \lambda^{-1}}{1 + 2 W \lambda^{-1}} \hat{P}. \label{eq:def_P_avg_cosine}
	\end{IEEEeqnarray}
	Note, that we cannot evaluate the actual SNR $\rho^{\ast}$ as we only obtain an upper and a lower bound on $\sigma^2_{\rv{\tilde{x}}}$, cf. Section~\ref{sec:LP-distortion}. Thus, we define the SNR w.r.t. $\rv{x}(t)$ as
	\begin{IEEEeqnarray}{rCL}
		\rho = \frac{P}{N_0 W} \ge \rho^{\ast}. \label{eq:SNR}
	\end{IEEEeqnarray}
%--------------------------------------------------------------------------------------------------------------------------------------------------------------------------------------------------------------------- %
\subsection{Error Events}
\label{subsec:System_Errors}
Transmitting the signal $\rv{x}(t)$ over the channel described in the previous section, including LP-distortion and AWGN, may cause three types of error events:
\begin{itemize}
	\item shifts of zero-crossings leading to errors in the magnitudes of the received symbol corresponding to $\rv{A}_k$
	\item insertion of zero-crossings causing an insertion of received symbols
	\item deletion of zero-crossing pairs, leading to the deletion of received symbols.
\end{itemize}

For channels with insertions and deletions are, to the best of our knowledge, only capacity bounds for binary channels available, e.g., \cite{gallager1961,Zig69,Fertonani2011,diggavi2007capacity}.
% --------------------------------------------------------------------------------------------------------------------------------------------------------------------------------------------------------------------- %
\subsection{Assumptions}
\label{subsec:System_Assumptions}
The following assumptions are made in the remainder of the paper in order to analyze the achievable rate. All these assumptions are reasonable and will be justified by argumentation and/or numerical simulations.
	\begin{enumerate}[leftmargin=1.0cm, label={(A\arabic*)}]
		\setItemnumber{1}
		\item \label{item:A1c}Throughout our analysis, the LP-distortion error $\rv{\tilde{x}}(t)$ is approximated to be Gaussian, which enables closed form analytical treatment.
	\end{enumerate}
We observe in simulations that this is valid for the parameter range of $W$ and $\lambda$ that maximizes the lower bound on the mutual information rate for a given bandwidth, see Appendix~\ref{app:distr_LP-distortion}. This results in a valid lower bound on the mutual information rate of the 1-bit quantized bandlimited AWGN-channel. Thus, we assume
	\begin{IEEEeqnarray}{rCL}
		\rv{z}(t) \sim \mathcal{N}(0,\sigma_\rv{z}^2). \label{eq:p_z(t)}
	\end{IEEEeqnarray}
\begin{enumerate}[leftmargin=1.0cm, label={(A\arabic*)}]
	\setItemnumber{2}
	\item \label{item:A2}For the considered input signals and the high SNR scenario, the occurrence of deletions is assumed to be negligible.  
\end{enumerate}
This is due to the facts that $\rv{A}_k \ge \beta$ and that $\beta$ depends directly on the bandwidth of the receiver, cf. (\ref{eq:Relation_W_beta}). Thus, the samples of the filtered noise with a temporal distance larger than $\beta$ can assumed to be uncorrelated and the possibility of a noise event inverting an entire symbol can be neglected. This argumentation has been verified by simulation, see Appendix~\ref{app:deletions}.

Hence, out of the error events described in Section~\ref{subsec:System_Errors}, the channel model between the random quantities $\Vrv{A}^{(K)}$ and $\Vrv{D}^{(M)}$ has to comprise magnitude errors caused by zero-crossing shifts, which we will denote by
\begin{IEEEeqnarray}{rCl}
	\Vrv{S}^{(K+1)} = [\rv{S}_0, \rv{S}_1, ... \rv{S}_K] \label{eq:S(K+1)_vector}
\end{IEEEeqnarray}
and insertion errors that we will denote by the process $\Vrv{V}^{(K)} = [\rv{V}_1,\rv{V}_2,...\rv{V}_K]$, both of which will be defined in Section~\ref{sec:error_events}.
\begin{enumerate}[leftmargin=1.0cm, label={(A\arabic*)}]
	\setItemnumber{3}
	\item \label{item:A4}There is only one zero-crossing in each transition interval $\left[\rv{T}_k-\frac{\beta}{2},\rv{T}_k+\frac{\beta}{2}\right]$.
\end{enumerate}
This follows from the bandlimitation of the noise, which prevents the signal from rapid changes, and it has been verified by numerical computation based on curve-crossing problems for Gaussian random processes. The results are presented in Appendix~\ref{app:ZC-count-Tk} and show that this assumption is fulfilled for an SNR above 5\,dB.
\begin{enumerate}[leftmargin=1.0cm, label={(A\arabic*)}]
	\setItemnumber{4}
	\item \label{item:A3}The individual elements of the processes $\Vrv{S}$ and $\Vrv{V}$ are i.i.d.
\end{enumerate}	
Due to \ref{item:A4} both error events can be separated in time: The shifting errors occur in the transition intervals $\left[\rv{T}_k-\sfrac{\beta}{2},\rv{T}_k+\sfrac{\beta}{2}\right]$ whereas the additional zero-crossings are of relevance during the hold time 
$\left[\rv{T}_k+\sfrac{\beta}{2},\rv{T}_{k+1}-\sfrac{\beta}{2}\right]$. 
This results in temporal separation of the individual $\rv{S}_k$, whose causative noise samples are spaced at least the minimum symbol duration $\beta$ apart. As $\beta$ is matched to the bandwidth of the noise, cf. (\ref{eq:Relation_W_beta}), the noise events causing the individual $\rv{S}_k$ and, hence, the individual $\rv{S}_k$ themselves are mutually independent. For the same reasoning the individual $\rv{V}_k$ are mutually independent as they are temporally separated by the transition time $\beta$, cf. (\ref{eq:f(t)_cosine}), leading as well to independent causative noise events.
\begin{enumerate}[leftmargin=1.0cm, label={(A\arabic*)}]
	\setItemnumber{5}			
	\item \label{item:A5}We focus on the mid to high SNR-domain and, thus, assume that the shifting errors $\rv{S}_{k} \ll \beta$. 
\end{enumerate}
	This approximation is valid for SNR values above 6\,dB, see Appendix~\ref{app:Gauss_Approx}.

% % % % % % % % % % % % % % % % % % % % % % % % % % % % % % % % % % % % % % % % % % % % % % % % % % % % % % % % % % % % % % % % % % % % % % % % % % % % % % % % % % % % % % % % % % % % % % % % % % % % % % % % % % % % %
% --------------------------------------------------------------------------------------------------------------------------------------------------------------------------------------------------------------------- %
% % % % % % % % % % % % % % % % % % % % % % % % % % % % % % % % % % % % % % % % % % % % % % % % % % % % % % % % % % % % % % % % % % % % % % % % % % % % % % % % % % % % % % % % % % % % % % % % % % % % % % % % % % % % %	
	
	\section{Bounding the Achievable Rate}
	\label{sec:error_events}	
	The capacity of a communication channel represents the highest rate at which we can transmit over the channel with an arbitrary small probability of error and is defined as
	\begin{IEEEeqnarray}{rCl}
	C = \sup~I'\left(\boldsymbol{\rv{A}};\boldsymbol{\rv{D}}\right) \label{eq:capacity}
	\end{IEEEeqnarray}
	where the supremum is taken over all distributions of the input signal, for which $\rv{\hat{x}}(t)$ is constrained to the average power $P$ and the bandwidth $W$. In (\ref{eq:capacity}) the mutual information rate is given by
	\begin{IEEEeqnarray}{rCl}
	I'\left(\boldsymbol{\rv{A}};\boldsymbol{\rv{D}}\right) = \lim\limits_{K \rightarrow \infty} \frac{1}{K T_{\text{avg}}}~I\left(\boldsymbol{\rv{A}}^{(K)};\boldsymbol{\rv{D}}^{(M)}\right) \label{eq:ach_rate}
	\end{IEEEeqnarray}
	with $I\big(\bm{\mathsf{A}}^{(K)};\bm{\mathsf{D}}^{(M)}\big)$ being the mutual information. Note that we have defined the mutual information rate based on a normalization with respect to the expected transmission time $K T_{\textrm{avg}}$. In the present paper, we derive a lower bound on the capacity by restricting ourselves to input signals as described in Section~\ref{subsec:System_Input}. However, later we will consider the supremum of $I'\big(\boldsymbol{\rv{A}};\boldsymbol{\rv{D}}\big)$ over the parameter $\lambda$ of the distribution of the $\rv{A}_k$ in (\ref{eq:p_A(a)}). Furthermore, we derive an upper bound on the achievable rate of this specific signaling scheme in order to quantify the impact of the bounding steps taken.
		
	The capacity with 1-bit quantization in (\ref{eq:capacity}) is always smaller than the AWGN capacity, which can be illustrated by the following calculation. The average number $\mu_0$ of zero-crossings of a Gaussian random process in a time interval of length $T$ is given by the Rice formula \cite{Rice1944}. For bandlimited Gaussian noise, it is
	$\mu_0~=~\sfrac{2}{\sqrt{3}} W T$.
	This corresponds to the average number of symbols per time interval $T$ in the 1-bit quantized continuous-time channel and is, by the factor $\sfrac{1}{\sqrt{3}} \approx 0.5774$, smaller than the number of $2 W T$  independent samples in an AWGN channel without amplitude quantization in the same time interval.
	
With \ref{item:A2} the error events to be considered are shifts and insertions of zero-crossing. 
Insertions are synchronization errors, that prevent the receiver from correctly identifying the beginning of a transmit symbol. Dobrushin has proven information stability and Shannon's coding theorem for channels with synchronization errors given discrete and finite random variables \cite{Dobrushin1967}. For the case of the continuous random processes $\Vrv{A}$ and $\Vrv{D}$ this proof remains for future work. However, we provide an intuition in Appendix~\ref{app:Coding_Theorem} why we consider it possible to design a code that can identify and correct insertions in this scenario and, thus, give an operational meaning to the mutual information rate we derive.

In order to analyze the achievable rate, we separately evaluate the impact of the shifted and inserted zero-crossings. For this purpose, we use the concept of a genie-aided receiver as in \cite{Fertonani2011}, which has information on inserted zero-crossings contained in an auxiliary process $\Vrv{V}$. Based on $\Vrv{V}$, which is described below, the genie-aided receiver
can remove the additional zero-crossings. Let $\boldsymbol{\rv{\hat{D}}}$ contain the temporal distances of the zero-crossings at the receiver when the additional zero-crossings are removed. The process $\boldsymbol{\rv{\hat{D}}}$ can be determined based on $\boldsymbol{\rv{{D}}}$ and $\boldsymbol{\rv{V}}$ such that the mutual information rate in case the receiver has side information about the inserted zero-crossings is given by
	\begin{IEEEeqnarray}{rCl}
		I'(\boldsymbol{\rv{A}};\boldsymbol{\hat{\rv{D}}}) = I'(\boldsymbol{\rv{A}};\boldsymbol{\rv{D}},\boldsymbol{\rv{V}}). \label{eq:I(A;D,V)=I(A;D_dach)}
	\end{IEEEeqnarray}
	Using the chain rule, we have
	\begin{align}
	I'(\boldsymbol{\rv{A}};\boldsymbol{\rv{D}},\boldsymbol{\rv{V}}) &= I'(\boldsymbol{\rv{A}};\boldsymbol{\rv{D}}) + I'(\boldsymbol{\rv{A}};\boldsymbol{\rv{V}}|\boldsymbol{\rv{D}}).
	\end{align}
	Thus
	\begin{align}
	I'(\boldsymbol{\rv{A}};\boldsymbol{\rv{D}}) &= I'(\boldsymbol{\rv{A}};\boldsymbol{\rv{D}},\boldsymbol{\rv{V}}) - I'(\boldsymbol{\rv{A}};\boldsymbol{\rv{V}}|\boldsymbol{\rv{D}}) \label{eq:Inf_rate_2c}
	\end{align}
	where $I'(\boldsymbol{\rv{A}};\boldsymbol{\rv{D}})$ is the mutual information rate without the side information on additional zero-crossings at the receiver. The effect of the shifted zero-crossings is captured in $I'(\boldsymbol{\rv{A}};\boldsymbol{\rv{D}},\boldsymbol{\rv{V}})$ and the influence of the inserted zero-crossings is described by $I'(\boldsymbol{\rv{A}};\boldsymbol{\rv{V}}|\boldsymbol{\rv{D}})$.
	Given that  the latter can only cause a degradation of the mutual information rate as mutual information is always non-negative, for the purpose of an upper bound on the mutual information rate $I'(\Vrv{A};\Vrv{D})$, it suffices to derive an upper bound on the mutual information rate $I'(\Vrv{A};\Vrv{\hat{D}})$ of the genie-aided receiver. I.e., independent of the nature of the auxiliary process $\boldsymbol{\rv{V}}$ it holds that
	\begin{IEEEeqnarray}{rCl}
		I'(\boldsymbol{\rv{A}};\boldsymbol{\rv{D}}) \le \bar{I}'(\boldsymbol{\rv{A}};\boldsymbol{\rv{D}}) = \bar{I}'(\boldsymbol{\rv{A}};\boldsymbol{\rv{D}},\boldsymbol{\rv{V}}). \label{eq:upperBound_1}
	\end{IEEEeqnarray}	
	For the characterization of the auxiliary process $\boldsymbol{\rv{V}}$, we consider for the moment the transmission of one input symbol $\rv{A}_k$. Its bounding zero-crossings $\rv{T}_{k-1}$ and $\rv{T}_k$ will be shifted to $\rv{\hat{T}}_{k-1}$ and $\rv{\hat{T}}_k$ by the noise process, such that
	\begin{IEEEeqnarray}{rCl}
		\rv{\hat{T}}_k = \rv{T}_k + \rv{S}_k \label{eq:Sk-Tk_relation}
	\end{IEEEeqnarray}
	where $\rv{S}_k$ is the $(k+1)$th time shift in $\Vrv{S}^{(K+1)}$, cf. (\ref{eq:S(K+1)_vector}).
	Additionally introduced zero-crossings will divide the input symbol $\rv{A}_k$ into a vector of corresponding received symbols. The latter is reversible, if the receiver knows which zero-crossings correspond to the originally transmitted ones. The receiver needs to sum up the distances $\rv{D}_m$ that are separated by the additional zero-crossings in order to obtain the corresponding symbols $\rv{\hat{D}}_k$. Intuitively, one would start such an algorithm with the first received symbol, which gives way to the following thought: Instead of providing the receiver with the exact positions in time of the additional zero-crossings, it suffices to know for each transmit symbol $\rv{A}_k$ how many received symbols have to be summed up to generate $\rv{\hat{D}}_k$ and, thus, obtain the sequence $\boldsymbol{\rv{\hat{D}}}^{(K)}$. Hence, the auxiliary sequence $\boldsymbol{\rv{V}}^{(K)}$ consists of positive integer numbers $\rv{V}_k \in \mathbb{N}$, representing for each input symbol the number of corresponding output symbols. Thus, the auxiliary process $\boldsymbol{\rv{V}}$ is discrete, which we use for lower-bounding the information rate in (\ref{eq:Inf_rate_2c}) by
	\begin{IEEEeqnarray}{rCl}
	I'(\boldsymbol{\rv{A}};\boldsymbol{\rv{D}}) && =  {I'(\boldsymbol{\rv{A}};\boldsymbol{\rv{D}},\boldsymbol{\rv{V}})} - H'(\boldsymbol{\rv{V}}|\boldsymbol{\rv{D}}) + H'(\boldsymbol{\rv{V}}|\boldsymbol{\rv{D}},\boldsymbol{\rv{A}}) \nonumber\\
	&& \ge I'(\boldsymbol{\rv{A}};\boldsymbol{\rv{D}},\boldsymbol{\rv{V}}) - {{H'(\boldsymbol{\rv{V}}|\boldsymbol{\rv{D}})}} \label{eq:lowerBound_0}\\
	&& \ge I'(\boldsymbol{\rv{A}};\boldsymbol{\rv{D}},\boldsymbol{\rv{V}}) - {H'(\boldsymbol{\rv{V}})} = 	\underline{I}'(\boldsymbol{\rv{A}};\boldsymbol{\rv{D}}) \label{eq:lowerBound_1}
	\end{IEEEeqnarray}
	where (\ref{eq:lowerBound_0}) results from the fact that the entropy rate of a discrete random process is non-negative and (\ref{eq:lowerBound_1}) is due to the fact that conditioning cannot increase entropy. 
	In the following, we will derive bounds on $I'(\boldsymbol{\rv{A}};\boldsymbol{\rv{D}},\boldsymbol{\rv{V}})$ and $H'(\boldsymbol{\rv{V}})$.

% % % % % % % % % % % % % % % % % % % % % % % % % % % % % % % % % % % % % % % % % % % % % % % % % % % % % % % % % % % % % % % % % % % % % % % % % % % % % % % % % % % % % % % % % % % % % % % % % % % % % % % % % % % % %
% --------------------------------------------------------------------------------------------------------------------------------------------------------------------------------------------------------------------- %
% % % % % % % % % % % % % % % % % % % % % % % % % % % % % % % % % % % % % % % % % % % % % % % % % % % % % % % % % % % % % % % % % % % % % % % % % % % % % % % % % % % % % % % % % % % % % % % % % % % % % % % % % % % % %	
\section{Achievable Rate of the Genie-Aided Receiver}
\label{sec:genie-aided-rx}	
	To evaluate the achievable rate $I'(\boldsymbol{\rv{A}};\boldsymbol{\rv{\hat{D}}})$ of the genie-aided receiver, cf. (\ref{eq:I(A;D,V)=I(A;D_dach)}), (\ref{eq:upperBound_1}) and (\ref{eq:lowerBound_1}), we have to evaluate the mutual information rate between the sequence of temporal spacings of zero-crossings at the channel input $\boldsymbol{\rv{A}}^{(K)}$ and sequence of the temporal spacing of zero-crossings of $\boldsymbol{\rv{\hat{D}}}^{(K)}$. Note, that in contrast to the original channel here both vectors $\boldsymbol{\rv{A}}^{(K)}$ and $\boldsymbol{\rv{\hat{D}}}^{(K)}$ are of same length as additional zero-crossings are removed at the receiver. The only error remaining is a shift $\rv{S}_k$ of every zero-crossings instant $\rv{T}_k$ to $\hat{\rv{T}}_k$. Hence, on a symbols level we can write with (\ref{eq:Sk-Tk_relation}) and (\ref{eq:Ak-Tk_relation}) for the channel output
	\begin{IEEEeqnarray}{rCl}
		\hat{\rv{D}}_k  = \hat{\rv{T}}_k - \hat{\rv{T}}_{k-1} 
				  = \rv{A}_k + \rv{S}_k - \rv{S}_{k-1} = \rv{A}_k + \rv{\Delta}_k.\label{eq:channel_output}
	\end{IEEEeqnarray}
	In order to derive an upper and lower bound on the mutual information rate of this channel, knowledge on the probability distribution of $\rv{S}_k$ is required.
	
% --------------------------------------------------------------------------------------------------------------------------------------------------------------------------------------------------------------------- %
	\subsection{The Distribution of the Shifting Errors}
	\label{subsec:genie-aided-rx_1}
	
	The distribution of $\mathsf{S}_{k}$ can be evaluated by mapping the probability density function (pdf) of the additive noise $\rv{z}(\mathsf{T}_{k})$ at the time instant $\mathsf{T}_{k}$ by the function 
	\begin{IEEEeqnarray}{rCL}
		\rv{z}(\mathsf{T}_{k})&=&-\sqrt{\hat{P}} f(\mathsf{S}_{k})=-\sqrt{\hat{P}}\sin\left(\frac{\pi}{\beta}\mathsf{S}_{k}\right)
	\end{IEEEeqnarray}
	into the zero-crossing error $\mathsf{S}_{k}$ on the time axis. The mapping hereby depends on the transition waveform $f(t)$.	
	{As $\rv{x}(t)$ is almost bandlimited, it can  be adequately described by a sampled representation with sampling rate $1/\beta$ to fulfill the Nyquist condition, cf.\ (\ref{eq:Relation_W_beta}).} Note that we here refer to the concept of sampling only to evaluate the value of $\rv{z}(t)$ at the time instant $\rv{T}_k$ of the original zero-crossing. We still assume the receiver to be able to resolve the zero-crossings instants with infinite resolution.
	With (\ref{eq:p_z(t)}) we have
	\begin{IEEEeqnarray}{rCL}
		&p_{\rv{S}}(s) =\left|\frac{\partial f(s)}{\partial s} p_{\rv{z}}(f(s))\right| 
		=\sqrt{\frac{\pi \hat{P}}{2\sigma_\rv{z}^2}}\frac{\cos\left(\frac{\pi}{\beta}s\right)}{\beta}\exp\left\{-\frac{\hat{P}}{2\sigma_\rv{z}^2}\sin^{2}\left(\frac{\pi}{\beta}s\right)\right\}.\label{PDF_S}
	\end{IEEEeqnarray}
	Given \ref{item:A5}, the pdf in (\ref{PDF_S}) can be well approximated by
	\begin{IEEEeqnarray}{rCL}
		p_S(s)&=&\sqrt{\frac{\pi \hat{P}}{2\sigma_\rv{z}^2}}\frac{1}{\beta}\exp\left\{-\frac{\hat{P}}{2\sigma_\rv{z}^2}\left(\frac{\pi}{\beta}s\right)^{2}\right\}. \label{eq:PDF_S_Gauss}
	\end{IEEEeqnarray}
	Hence, in the high SNR case the zero-crossing errors $\mathsf{S}_{k}$ are approximately Gaussian distributed, i.e., $\mathsf{S}_{k}\sim\mathcal{N}\left(0,\sigma_{\mathsf{S}}^{2}\right)$ with
	\begin{IEEEeqnarray}{rCL}
		\sigma_{\mathsf{S}}^{2}&=&\frac{\sigma_\rv{z}^2}{4\pi^{2}W^{2}\hat{P}}.\label{eq:sigmaS_Def}
	\end{IEEEeqnarray}
% --------------------------------------------------------------------------------------------------------------------------------------------------------------------------------------------------------------------- %
	\subsection{Upper Bound on the Achievable Rate of the Genie-Aided Receiver}
	\label{subsec:genie-aided-rx_3}
	Given $\mathsf{S}_{k}\sim\mathcal{N}\left(0,\sigma_{\mathsf{S}}^{2}\right)$, cf. (\ref{eq:sigmaS_Def}), we have for $\rv{\Delta}_k = \rv{S}_k - \rv{S}_{k-1}$ in (\ref{eq:channel_output})
	\begin{IEEEeqnarray}{rCL}
		\rv{\Delta}_k \sim \mathcal{N}(0,2\sigma_\rv{S}^2) \label{eq:p_Delta_k}
	\end{IEEEeqnarray}
	as the $\rv{S}_k$ are independent, see \ref{item:A3}.
	The values $\rv{\Delta}_k$ are correlated as they always depend on the current and the previous $\rv{S}_k$, such that the autocorrelation function (ACF) of $\Vrv{\Delta}$ is given by
	\begin{IEEEeqnarray}{rCL}
		\phi_{\Delta \Delta}(k) = \E\left[\Delta_l \Delta_{l+k}\right] = \begin{cases}
			2 \sigma_s^2, & k=0\\ - \sigma_s^2, &|k|=1\\ 0, & \text{otherwise}
		\end{cases} \label{eq:ACF_Delta}
	\end{IEEEeqnarray}
	which yields for the covariance matrix $\mathbf{R}^{(K)}_{\rv{\Delta}}$ of zero-crossing shifts $\Vrv{\Delta}^{(K)}=[\rv{\Delta}_{1},...,\rv{\Delta}_K]$
	\begin{IEEEeqnarray}{rCL}
		\mathbf{R}^{(K)}_{\rv{\Delta}} = \E\left[\Vrv{\Delta}^{(K)^T} \Vrv{\Delta}^{(K)}\right]
		&=&\sigma_{\mathsf{S}}^{2}\left(\begin{smallmatrix}
			2 & -1 & 0 & \hdots & 0\\
			-1 & 2 & -1 & \ddots & \vdots\\
			0 & -1 & 2 & \ddots & 0\\
			\vdots & \ddots & \ddots & \ddots & -1\\
			0 & \hdots & 0 & -1 & 2
		\end{smallmatrix}\right). \label{eq:Cov_mtrx_Delta}
	\end{IEEEeqnarray}
	Hence, the channel with the genie-aided receiver is a colored additive Gaussian noise channel with input $\Vrv{A}$, output $\Vrv{\hat{D}}$, and noise $\Vrv{\Delta}$. The capacity of the colored additive Gaussian noise channel is achieved for Gaussian distributed input symbols \cite[Chapter 9, Eq. (9.97)]{CoverBook2} and provides an upper bound on the mutual information rate of the channel with the genie-aided receiver. Thus, we get
	\begin{IEEEeqnarray}{rCL}
		I'(\boldsymbol{\rv{A}};\boldsymbol{\rv{D}},\boldsymbol{\rv{V}}) &&\le \bar{I}'(\boldsymbol{\rv{A}};\boldsymbol{\rv{D}},\boldsymbol{\rv{V}}) 
		= \frac12 \int_{-\frac12}^{\frac12} \log \left(1+\frac{(\nu-S_\rv{\Delta}(f))^{+}}{S_\rv{\Delta}(f)}\right) df \label{eq:upper_bound_ach_rate}
	\end{IEEEeqnarray}
	where $\nu$ is chosen such that  
	\begin{IEEEeqnarray}{rCL}
		\int_{-\frac12}^{\frac12} (\nu-S_{\rv{\Delta}}(f))^{+} df = \sigma_\rv{A}^2 \label{eq:cond_waterlevel}
	\end{IEEEeqnarray}
	with $\sigma_\rv{A}^2=\sfrac{1}{\lambda^2}$ being the variance of the $\rv{A}_k$, cf. (\ref{eq:input_variance}). Moreover, $S_\rv{\Delta}(f)$ is the PSD of $\Vrv{\Delta}$ and it is given by the $z$-transform of (\ref{eq:ACF_Delta})
	as
	\begin{IEEEeqnarray}{rCL}
		\label{eq:col_noise_spectrum}
		S_\rv{\Delta}(f)=2 \sigma_\rv{S}^2 (1-\cos(2 \pi f)), \qquad |f|<0.5. \label{eq:S_Delta(f)}
	\end{IEEEeqnarray}
	Although $S_\rv{\Delta}(f)$ is equal to zero for $f = 0$ it can be shown that the integral in (\ref{eq:upper_bound_ach_rate}) exists, using that $\nu \ge (\nu-S_{\rv{\Delta}}(f))^{+}~\forall f$ and solving
	\begin{IEEEeqnarray}{rCl}
		\int_{-\frac{1}{2}}^{\frac{1}{2}} \log\left(1 +\frac{a}{1-\cos(2 \pi f)}\right) df = \arcosh\left(a+1\right) \label{eq:Integral}
	\end{IEEEeqnarray}
	where $a=\sfrac{\nu}{(2 \sigma^2_\rv{S})} \in \mathbb{R}$ and $a > 0$.
		
	\subsection{Lower Bound on the Achievable Rate of the Genie-Aided Receiver}
	\label{subsec:genie-aided-rx_2}

	The mutual information between the temporal spacings of the zero-crossings of the channel input signal $\bm{\mathsf{A}}^{(K)}$ on the one hand, and the zero-crossings of the signal at the input to the 1-bit quantizer $\bm{\rv{\hat{D}}}^{(K)}$  on the other hand is given by
\begin{IEEEeqnarray}{rCL}
	I\big(\bm{\mathsf{A}}^{(K)};\bm{\mathsf{\hat{D}}}^{(K)}\big)&&=h\big(\bm{\mathsf{A}}^{(K)}\big)-h\big(\bm{\mathsf{A}}^{(K)}|\bm{\mathsf{\hat{D}}}^{(K)}\big)\nonumber\\
	&&=h\big(\bm{\mathsf{A}}^{(K)}\big)-h\big(\bm{\mathsf{A}}^{(K)}-\hat{\bm{\mathsf{A}}}_{\textrm{LMMSE}}^{(K)}\big|\bm{\mathsf{\hat{D}}}^{(K)}\big)\label{MutInfDiffEntropSep}
\end{IEEEeqnarray}
	where $h(\cdot)$ denotes the differential entropy. Moreover, $\hat{\bm{\mathsf{A}}}_{\textrm{LMMSE}}^{(K)}$ is the linear minimum mean-squared error estimate of $\bm{\mathsf{A}}^{(K)}$ based on $\bm{\rv{\hat{D}}}^{(K)}$. Equality (\ref{MutInfDiffEntropSep}) follows from the fact that addition of a constant does not change differential entropy and the fact that $\hat{\bm{\mathsf{A}}}_{\textrm{LMMSE}}^{(K)}$ can be treated as a constant while conditioning on $\bm{\rv{\hat{D}}}^{(K)}$ as it is a deterministic function of $\bm{\rv{\hat{D}}}^{(K)}$.
	
	Next, we will upper-bound the second term on the RHS of (\ref{MutInfDiffEntropSep}), i.e., $h\big(\bm{\mathsf{A}}^{(K)}-\hat{\bm{\mathsf{A}}}_{\textrm{LMMSE}}^{(K)}\big|\bm{\rv{\hat{D}}}^{(K)}\big)$. This term describes the randomness of the linear minimum mean-squared estimation error while estimating $\bm{\mathsf{A}}^{(K)}$ based on the observation $\bm{\rv{\hat{D}}}^{(K)}$. It can be upper-bounded by the differential entropy of a Gaussian random variable having the same covariance matrix \cite[Theorem~8.6.5]{CoverBook2}. The estimation error covariance matrix of the linear minimum mean-squared error (LMMSE) estimator is given by
	\begin{IEEEeqnarray}{rCL}
		\mathbf{R}_{\textrm{err}}^{(K)}
		&=&\E\big[
		\big(\bm{\mathsf{A}}^{(K)}-\hat{\bm{\mathsf{A}}}_{\textrm{LMMSE}}^{(K)}\big)\big(\bm{\mathsf{A}}^{(K)}-\hat{\bm{\mathsf{A}}}_{\textrm{LMMSE}}^{(K)}\big)^{T}\big]\nonumber\\
		&=&\sigma_{\mathsf{A}}^{2}\mathbf{I}^{(K)}-\sigma_{\mathsf{A}}^{4}\big(\sigma_{\mathsf{A}}^{2}\mathbf{I}^{(K)}+\mathbf{R}_{\rv{\Delta}}^{(K)}\big)^{-1}\label{eq:ErrEstCov}
	\end{IEEEeqnarray}
	where $\sigma_{\rv{A}}^2$ is given in (\ref{eq:input_variance}). 	
	Furthermore, $\mathbf{I}^{(K)}$ is the identity matrix of size $K\times K$ and $\mathbf{R}_{\mathsf{\Delta}}^{(K)}$ is the covariance matrix of the shifting error $\Vrv{\Delta}^{(K)}$ and given in (\ref{eq:Cov_mtrx_Delta}).
	Thus, the differential entropy $h(\bm{\mathsf{A}}^{(K)}-\hat{\bm{\mathsf{A}}}_{\textrm{LMMSE}}^{(K)}\big|\bm{\rv{\hat{D}}}^{(K)})$ is upper-bounded by
	\begin{IEEEeqnarray}{rCL}
		h(\bm{\mathsf{A}}^{(K)}-\hat{\bm{\mathsf{A}}}_{\textrm{LMMSE}}^{(K)}\big|\bm{\mathsf{D}}^{(K)})&\le&\frac{1}{2}\log\det\left(2\pi e \mathbf{R}_{\textrm{err}}^{(K)}\right)\label{UppBoundDiffEntropCond}
	\end{IEEEeqnarray}
	yielding the following lower bound for the mutual information in (\ref{MutInfDiffEntropSep})
	\begin{IEEEeqnarray}{rCL}
		I(\bm{\mathsf{A}}^{(K)};\bm{\rv{\hat{D}}}^{(K)}) &&\ge h(\bm{\mathsf{A}}^{(K)})-\frac{1}{2}\log\det\left(2\pi e \mathbf{R}_{\textrm{err}}^{(K)}\right)\nonumber\\
		&&=K h(\mathsf{A}_{k})+\frac{1}{2}\log\det\!\left(\!(2\pi e)^{\!-1}\! 
		\left(\sigma_{\mathsf{A}}^{-2}\mathbf{I}^{(K)}\!+\!(\mathbf{R}_{\mathsf{\Delta}}^{(K)})^{-1}\!\right)\!
		\right)\label{LowBouw2}
	\end{IEEEeqnarray}
	where the first term of (\ref{LowBouw2}) follows from the independence of the elements of $\bm{\mathsf{A}}^{(K)}$ and for the second term we have used (\ref{eq:ErrEstCov}) and the matrix inversion lemma.
	With (\ref{LowBouw2}) the mutual information rate in (\ref{eq:I(A;D,V)=I(A;D_dach)}) is lower-bounded by
	\begin{IEEEeqnarray}{rCL}
		I'(\bm{\mathsf{A}};\bm{\rv{\hat{D}}})	&&\ge \lim_{K\rightarrow\infty}\frac{1}{KT_{\textrm{avg}}}\bigg\{
		K h(\mathsf{A}_{k}) 
		+\frac{1}{2}\log\det\left((2\pi e)^{-1} \left(\sigma_{\mathsf{A}}^{-2}\mathbf{I}^{(K)}+(\mathbf{R}_{\mathsf{\Delta}}^{(K)})^{-1}\right)\right)\bigg\} \nonumber\\
		&&=\frac{1}{T_{\textrm{avg}}}\bigg\{h(\mathsf{A}_{k}) 
		+\frac{1}{2}\int_{-\frac{1}{2}}^{\frac{1}{2}}\log\left(\frac{\sigma_{\rv{A}}^{-2}}{2\pi e} \left(1+\frac{\sigma_{\rv{A}}^2}{S_{\mathsf{\Delta}}(f)}\right)\right)df\bigg\}\label{LowBouw3}
	\end{IEEEeqnarray}
	where for (\ref{LowBouw3}) we have used Szeg\"o's theorem on the asymptotic eigenvalue distribution of Hermitian Toeplitz matrices \cite[pp.\ 64-65]{Szgo58}, \cite{Gray_ToeplitzReview}. Here, $S_{\mathsf{\Delta}}(f)$ is the PSD of $\Vrv{\Delta}$ given in (\ref{eq:S_Delta(f)}) and corresponding to the sequence of covariance matrices $\mathbf{R}_{\mathsf{\Delta}}^{(K)}$.
	Despite the discontinuity of the integrand in (\ref{LowBouw3}), it can be shown that the integral exists analogously as in (\ref{eq:Integral}), here with  $a=\sfrac{\sigma_{\rv{A}}^2}{(2 \sigma^2_\rv{S})}$.
	As $\mathsf{A}_{k}$ is exponentially distributed, we get
	\begin{IEEEeqnarray}{rCL}
		h(\mathsf{A}_{k})&=&1-\log(\lambda).\label{EntropExp}
	\end{IEEEeqnarray}
	With (\ref{eq:T_avg}), (\ref{eq:input_variance}), (\ref{eq:Relation_W_beta}), (\ref{eq:sigmaS_Def}), (\ref{eq:Integral}), and (\ref{EntropExp}), the lower bound in (\ref{LowBouw3}) can be written as
	\begin{IEEEeqnarray}{rCL}
		I'(\bm{\mathsf{A}};\bm{\rv{\hat{D}}}) && \ge\frac{1}{2 T_{\text{avg}}} \left\{\log\!\left(\frac{e}{2\pi}\right)+\arcosh\left(\frac{1}{2 \sigma_{\rv{S}}^2 \lambda^{2}}+1\right)\right\}\nonumber\\
		&&=\frac{W}{1+2W\lambda^{-1}}\left\{\log\!\left(\frac{e}{2\pi}\right) + \arcosh\left(\frac{2 \pi^2 W^2 \hat{P}}{\sigma_{\rv{z}}^2 \lambda^2}+1\right)\right\}.\label{Rate_LB}
	\end{IEEEeqnarray}

% % % % % % % % % % % % % % % % % % % % % % % % % % % % % % % % % % % % % % % % % % % % % % % % % % % % % % % % % % % % % % % % % % % % % % % % % % % % % % % % % % % % % % % % % % % % % % % % % % % % % % % % % % % % %
% --------------------------------------------------------------------------------------------------------------------------------------------------------------------------------------------------------------------- %
% % % % % % % % % % % % % % % % % % % % % % % % % % % % % % % % % % % % % % % % % % % % % % % % % % % % % % % % % % % % % % % % % % % % % % % % % % % % % % % % % % % % % % % % % % % % % % % % % % % % % % % % % % % % %	

\section{Characterization of the Process of Additional Zero-Crossings}
\label{sec:Aux_Proc}

In order to lower-bound the $I'(\boldsymbol{\rv{A}};\boldsymbol{\rv{D}})$, i.e., the rate without the side information provided by $\boldsymbol{\rv{V}}$ to the receiver, it remains to find an explicit expression or an upper bound for $H'(\boldsymbol{\rv{V}})$, cf. (\ref{eq:lowerBound_1}). For every input symbol $\rv{A}_k$ the random variable $\rv{V}_k$, which describes the number of received symbols that correspond to the transmitted one, depends on the number $\rv{N}_k$ of inserted zero-crossings by\looseness-1
	\begin{IEEEeqnarray}{rCl}
	\rv{V}_k = \rv{N}_k+1.
	\end{IEEEeqnarray}
	Hence, we need to determine the number of times within one symbol at which $\rv{z}(t) = -\rv{x}(t)$ as in this case the received signal will be zero. Based on assumption \ref{item:A4} we do not need to consider the transition intervals as they just contain the shifted zero-crossing. It remains the time $T_{\text{sat}} = \E[\rv{A}_k]-\beta = \lambda^{-1}$ in which the signal level $\pm \sqrt{\hat{P}}$ is maintained, leading to a level-crossing problem. Level-crossing problems, especially for Gaussian processes, have been studied over decades, e.g., by Kac \cite{kac1943}, Rice \cite{Rice1944}, Cramer and Leadbetter \cite{Cramer1967}.  
	In order to be able to derive a closed-form expression for the lower bound on $I'(\boldsymbol{\rv{A}};\boldsymbol{\rv{D}})$, we will derive an upper bound on $H'(\boldsymbol{\rv{V}})$ based on the first moment of the distribution of $\rv{V}_k$.
	For a stationary zero-mean Gaussian random process, the expected number of crossings of the level $\sqrt{\hat{P}}$ in the time interval $T_{\text{sat}} = \lambda^{-1}$  is given by the Rice formula \cite{Rice1944}
	\begin{IEEEeqnarray}{rCl}
	\mu = \E[\rv{V}_k] = \mathbb{E}[\rv{N}_k] +1 = \frac{1}{\pi} \sqrt{\frac{-s''_{\rv{zz}}(0)}{\sigma_\rv{z}^2}} \exp \left(- \frac{\hat{P}}{2 \sigma_\rv{z}^2}\right) \lambda^{-1} +1. \label{eq:mu}
	\end{IEEEeqnarray}
	Here, $s_{\rv{zz}}(\tau)$ is the ACF of the Gaussian process $\rv{z}(t)$ and $s''_{\rv{zz}}(\tau)=\sfrac{\partial}{\partial \tau^2} s_{\rv{zz}}(\tau)$. In order to ensure $\E[\rv{N}_k]$ to be finite, $-s''_{\rv{zz}}(0) < \infty$ has to hold. 
	Analogously to (\ref{eq:var_z(t)}), we have
	\begin{IEEEeqnarray}{rCl}
	s''_{\rv{zz}}(0) = s''_{\rv{\hat{n}\hat{n}}}(0) + s''_{\rv{\tilde{x}\tilde{x}}}(0) 
	 = -\frac43 N_0 W^3 + s''_{\rv{\tilde{x}\tilde{x}}}(0). \label{eq:def_s_zz(0)}
	\end{IEEEeqnarray}
	where $s''_{\rv{\tilde{x}\tilde{x}}}(0)$ is finite for finite bandwidths $W$, see Section \ref{sec:LP-distortion}.
	Using (\ref{eq:mu}) we upper-bound the entropy rate $H'(\boldsymbol{\rv{V}})$. For a given mean $\mu$, the entropy maximizing distribution for a positive, discrete random variable is the geometric distribution, cf. \cite[Section 2.1]{kapur1989maximum}. Hence, we can upper-bound the entropy $H(\rv{V}_k)$ by
	\begin{IEEEeqnarray}{rCl}
		H(\rv{V}_k) \le (1-\mu) \log \left(\mu - 1\right) + \mu \log \mu \label{eq:H(Vk)}.
	\end{IEEEeqnarray}
	The derivation of (\ref{eq:H(Vk)}) is given in Appendix~\ref{app:deriv-H(Vk)}. With \ref{item:A3} we obtain for the entropy rate of the auxiliary process
	\begin{IEEEeqnarray}{rCl}
		H'(\boldsymbol{\rv{V}}) = \frac{1}{T_{\text{avg}}} H(\rv{V}_k). \label{eq:entropy_rate_V}
	\end{IEEEeqnarray}
	Note that the bound on $H(\rv{V}_k)$ is an increasing function in $\mu$ and the expected number of level-crossings of the random Gaussian process increases with its variance $\sigma_\rv{z}^2$. Hence, to evaluate (\ref{eq:mu}), an upper bound for $\sigma_\rv{z}^2$ and, thus, for $\sigma_{\rv{\tilde{x}}}^2$ is required. An upper bound on $\sigma_{\rv{\tilde{x}}}^2$ results in a lower bound on $s''_{\rv{\tilde{x}\tilde{x}}}(0)$, cf. Section~\ref{subsec:Lowpass_B}, as the two parameters depend on the ACF of the distortion process $\rv{\tilde{x}}(t)$ and cannot be chosen independently. Both bounds  will be derived in the next section.
% % % % % % % % % % % % % % % % % % % % % % % % % % % % % % % % % % % % % % % % % % % % % % % % % % % % % % % % % % % % % % % % % % % % % % % % % % % % % % % % % % % % % % % % % % % % % % % % % % % % % % % % % % % % %
% --------------------------------------------------------------------------------------------------------------------------------------------------------------------------------------------------------------------- %
% % % % % % % % % % % % % % % % % % % % % % % % % % % % % % % % % % % % % % % % % % % % % % % % % % % % % % % % % % % % % % % % % % % % % % % % % % % % % % % % % % % % % % % % % % % % % % % % % % % % % % % % % % % % %	

\section{Signal Distortion by Lowpass-Filtering}
\label{sec:LP-distortion}
	The distortion of $\rv{x}(t)$ introduced by the lowpass-filter can be quantified by the clipped energy,
 using the mean squared error $\sigma^2_{\rv{\tilde{x}}}$ as distortion measure, which is given by
	\begin{IEEEeqnarray}{rCl}
	 \sigma^2_{\rv{\tilde{x}}} 
	 = \lim\limits_{T \rightarrow \infty} \frac{1}{T} \int_{-T}^{T} \E\left[\rv{\tilde{x}}^2(t)\right] dt
	  =  \frac{1}{2 \pi} \int_{-\infty}^{\infty} S_{\rv{\tilde{X}}}(\omega) d \omega \label{eq:def_mse}
	\end{IEEEeqnarray}
	where (\ref{eq:def_mse}) is Parseval's Theorem with $S_{\rv{\tilde{X}}}(\omega)$ being the PSD of $\rv{\tilde{x}}(t)$.
	As we consider a rectangular filter with cutoff-frequency $W$, it holds
	\begin{IEEEeqnarray}{rCl}
		S_{\rv{\tilde{X}}}(f) = \begin{cases}
		S_{\rv{X}}(f) & |f| > W\\
		0 & |f| \le W	
		\end{cases}
	\end{IEEEeqnarray}
	with $S_{\rv{X}}(f)$ being the PSD of $\rv{x}(t)$.
	As $S_\rv{X}(\omega)$ is even, we get
	\begin{IEEEeqnarray}{rCl}
			\sigma^2_{\rv{\tilde{x}}} = \frac{1}{\pi} \int_{2 \pi W}^{\infty}  S_{\rv{X}}(\omega) d \omega \label{eq:MSE(Sx(f))}
	\end{IEEEeqnarray}
	In order to evaluate (\ref{eq:MSE(Sx(f))}), we derive $S_\rv{X}(\omega)$. Steps on bounding $\sigma_{\rv{\tilde{x}}}^2$ and $s''_{\rv{\tilde{x}\tilde{x}}}(0)$ are taken subsequently.
% --------------------------------------------------------------------------------------------------------------------------------------------------------------------------------------------------------------------- %
	 \subsection{Signal Spectrum}
	 \label{subsec:Lowpass_A}
\begin{figure}
		\centering
		\includegraphics[width= 0.8\textwidth]{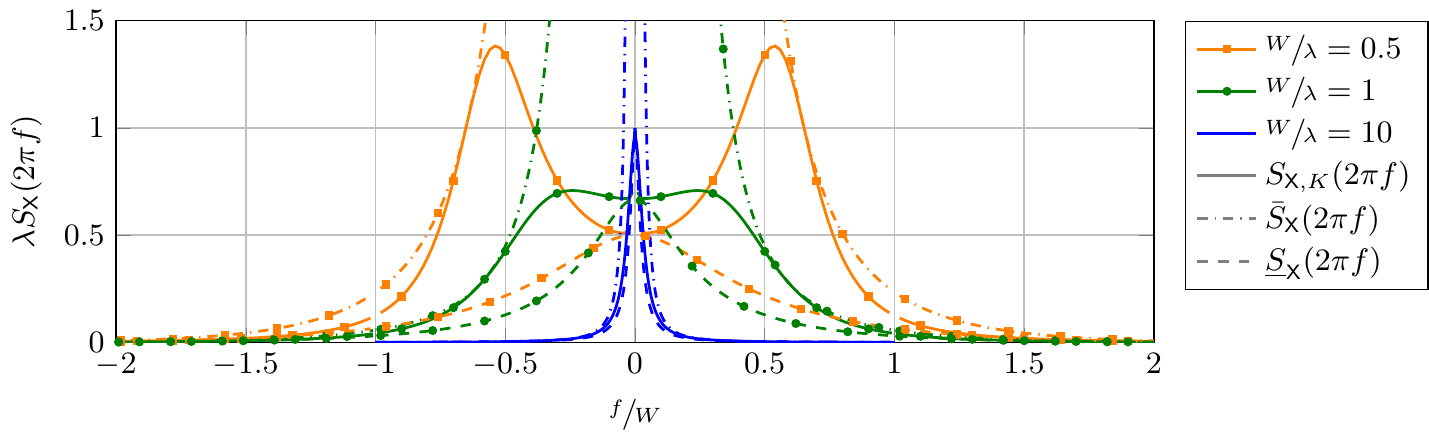}
		\caption{Upper bound on the PSD of $\rv{x}(t)$ for the sine-waveform (\ref{eq:f(t)_cosine})}
		\label{fig:S_X(w)}
	\end{figure}

	 The PSD of a random process is defined as 
	 \begin{IEEEeqnarray}{rCl}
	 		S_\rv{X}(\omega) = \lim\limits_{K \rightarrow \infty} \frac{\mathbb{E} \left[\left|\rv{X}(\omega)\right|^2\right]}{K T_{\text{avg}}} \label{eq:def_S(w)}
	 \end{IEEEeqnarray}
	 where $\rv{X}(\omega)$ is the spectrum of the random process $\rv{x}(t)$ defined in (\ref{eq:x(t)}) and given by 
	 \begin{IEEEeqnarray}{rCl}
	 		\rv{X}(\omega) = \left(\sum_{k=1}^{K} \sqrt{\hat{P}} (-1)^k \, G(\omega) \, e^{-j \omega \rv{T}_k} \right) + \sqrt{\hat{P}} 2 \pi \delta(\omega) \label{eq:S(x)}
	 \end{IEEEeqnarray}
	 where $G(\omega)$ is the Fourier transformation of the waveform $g(t)$ in (\ref{eq:g(t)_allgemein}).
	It holds that
	\begin{align}
		G(\omega) = -j \left[\frac{1+e^{-j\omega\beta}}{\omega}+e^{-j\omega \frac{\beta}{2}} a(\omega)\right] \label{eq:G(omega)}
	\end{align}
	where $a(\omega)$ is a real function in $\mathbb{R}$ given by
	\begin{IEEEeqnarray}{rCl}
		a(\omega) = - \frac{1}{j} \int_{-\frac{\beta}{2}}^{\frac{\beta}{2}} f(t) e^{-j \omega t} dt.
	\end{IEEEeqnarray}
	Based on this in  Appendix~\ref{app:spectrum_formula}, we show that the PSD of $\rv{x}(t)$ is given by 
	\begin{IEEEeqnarray}{rCl}
		S_\rv{X}(\omega) = \frac{\hat{P} \left|G(\omega)\right|^2}{T_{\text{avg}}} \left(1 + 2 \lim\limits_{K\rightarrow \infty}  \sum_{n=1}^{K-1} (-1)^n \left(1-\frac{n}{K}\right) \E[\cos(\omega \rv{L}_n)]\right) \label{eq:S_X(w)_with-k}
	\end{IEEEeqnarray}
	with $\left|G(\omega)\right|^2$ given in (\ref{eq:F(w)_App}) and where $n=k-j$ is the index describing the distance between two arbitrary zero-crossing instances and $\rv{L}_n = \rv{T}_k-\rv{T}_j$ is the corresponding random variable with probability distribution
	\begin{IEEEeqnarray}{rCl}
		p_\rv{L}(l_n) = \frac{\lambda^n e^{-\lambda \left(l_n - n \beta\right)} \left(l_n-n \beta\right)^{n-1}}{(n-1)!}, ~ n \ge 1, l_n \ge n \beta.  \label{eq:p(Xn)}
	\end{IEEEeqnarray}
	Using (\ref{eq:p(Xn)}) to calculate the expectation in (\ref{eq:S_X(w)_with-k}) yields
	\begin{IEEEeqnarray}{rCl}
		\mathbb{E}[\cos(\omega \rv{L}_n)]
		= \left(\frac{\lambda}{\sqrt{\lambda^2+\omega^2}}\right)^n \cos \left(n\left(\omega \beta + \arctan \left(\frac{\omega}{\lambda}\right)\right)\right)
		\le \left(\frac{\lambda}{\sqrt{\lambda^2+\omega^2}}\right)^n \label{eq:Expect_cos_omega_L}
	\end{IEEEeqnarray}
	which can be used to upper-bound the infinite sum in (\ref{eq:S_X(w)_with-k}) by
	\begin{IEEEeqnarray}{rCl}
		\lim\limits_{K \rightarrow \infty} \sum\limits_{n=1}^{K-1} \left(1-\frac{n}{K}\right) \left(\frac{\lambda}{\sqrt{\lambda^2+\omega^2}}\right)^n = \frac{\lambda}{\sqrt{\lambda^2+\omega^2} - \lambda} = c(\omega). \label{eq:c(omega)}
	\end{IEEEeqnarray}
	Hence, the PSD can be bounded as
	\begin{IEEEeqnarray}{rCl}
	S_\rv{X}(\omega) \le \frac{  \hat{P} }{T_{\text{avg}}}  (1+2 c(\omega)) \left|G(\omega)\right|^2 = \bar{S}_\rv{X}(\omega). \label{eq:S_X(w)_bound}
	\end{IEEEeqnarray}
	Numerically we find that with (\ref{eq:Expect_cos_omega_L}) the infinite sum in (\ref{eq:S_X(w)_with-k}) has periodic minima. They occur when $\left(\omega \beta + \arctan \left(\frac{\omega}{\lambda}\right)\right) = 2 m \pi$, $m \in \mathbb{Z}$, for which the cosine is always one such that it remains	 
		\begin{IEEEeqnarray}{rCl}
			\lim\limits_{K \rightarrow \infty} \sum\limits_{n=1}^{K-1} (-1)^n \left(1-\frac{n}{K}\right) \left(\frac{\lambda}{\sqrt{\lambda^2+\omega^2}}\right)^n = -\frac{\lambda}{\sqrt{\lambda^2+\omega^2} + \lambda}. \label{eq:lowerbound_inf_sum}
		\end{IEEEeqnarray}
		Based on (\ref{eq:lowerbound_inf_sum}) $S_\rv{X}(\omega)$ in (\ref{eq:S_X(w)_with-k}) can be lower bounded by
	\begin{IEEEeqnarray}{rCl}
		S_\rv{X}(\omega) \ge \frac{  \hat{P} }{T_{\text{avg}}}  \frac{\left|G(\omega)\right|^2}{(1+2 c(\omega))} = \underline{S}_\rv{X}(\omega) . \label{eq:S_X(w)_Lowerbound}
	\end{IEEEeqnarray}
		where we have used that $1-2 \frac{\lambda}{\sqrt{\lambda^2+\omega^2} + \lambda} = \frac{\sqrt{\lambda^2+\omega^2}-\lambda}{\sqrt{\lambda^2+\omega^2}+\lambda} = \frac{1}{1+2 c(\omega)}$ with $c(\omega)$ given in (\ref{eq:c(omega)}).
	For the sine-waveform introduced in (\ref{eq:f(t)_cosine}), we have
	\begin{IEEEeqnarray}{rCl}
		\left|G(\omega)\right|^2 = 2 (1+\cos(\omega \beta)) \left[\frac{\pi^2}{\omega(\pi^2-\omega^2 \beta^2)}\right]^2. \label{eq:F(omega)_cosine}
	\end{IEEEeqnarray}
	Fig.~\ref{fig:S_X(w)} shows an approximation of the normalized PSD $\lambda {S}_{\rv{X},K} \left(2 \pi f\right)$ for a finite $K=10^5$ as well as the upper bound on the spectrum for different ratios $\sfrac{W}{\lambda}$. As the abscissa is normalized on the one-sided bandwidth $W$, it can be seen that with increasing $\sfrac{W}{\lambda}$, the available bandwidth $2 W$ becomes less utilized. This has an impact on the spectral efficiency as will be discussed in Section~\ref{sec:Conclusion}.

% -------------------------------------------------------------------------------------------------------------------------------------------------------------------------------------------------------------------- %	
	\subsection{Bounds on $\sigma_\rv{\tilde{x}}^2$ and $s''_{\rv{\tilde{x}\tilde{x}}}(0)$}
	\label{subsec:Lowpass_B}
	With the results from Section \ref{subsec:Lowpass_A} bounds on $\sigma^2_{\rv{\tilde{x}}}$ and $s''_{\rv{\tilde{x}\tilde{x}}}(0)$ can be computed. For the upper bound on $\sigma^2_{\rv{\tilde{x}}}$ we get with (\ref{eq:MSE(Sx(f))}) and (\ref{eq:S_X(w)_bound}) 
	\begin{align}
		\sigma^2_{\rv{\tilde{x}}} &\le \frac{1}{\pi}  \int\limits_{2 \pi W}^{\infty} \frac{  \hat{P} }{T_{\text{avg}}}  (1+2 c(\omega)) \left|G(\omega)\right|^2 d\omega %\\	&
		 \le \frac{1}{\pi}  \frac{  \hat{P} }{T_{\text{avg}}}  (1+2 c_1) \int\limits_{2 \pi W}^{\infty} \left|G(\omega)\right|^2 d\omega. \label{eq:sigma_x_tilde_2_bounded}
	\end{align}
	In order to obtain (\ref{eq:sigma_x_tilde_2_bounded}) one further bounding step is applied. Note, that $c(\omega)$ is monotonically decreasing w.r.t. $|\omega|$ and, hence, for all $|\omega| \ge 2\pi W$
	\begin{IEEEeqnarray}{rCl}
		 c(\omega) \le c(2 \pi W) = \frac{\lambda}{\sqrt{\lambda^2+4 \pi^2 W^2}-\lambda} = c_1 \label{eq:def_c1}.
	\end{IEEEeqnarray}
	By using (\ref{eq:sigma_x_tilde_2_bounded}) for the upper bound and (\ref{eq:MSE(Sx(f))})  and (\ref{eq:S_X(w)_Lowerbound}) for the lower bound, respectively, we obtain the bounds on $\sigma_\rv{\tilde{x}}^2$. Additionally, we use (\ref{eq:F(omega)_cosine}) for the sine-waveform in (\ref{eq:f(t)_cosine}) and get
	\begin{align}
		\underline{\sigma}_{\rv{\tilde{x}}}^2 = \frac{\hat{P} \beta}{2 (1+2 c_1) T_{\text{avg}} \pi^2} c_0 &\le \sigma_{\rv{\tilde{x}}}^2  \le \frac{(1+2 c_1) \hat{P} \beta}{2 T_{\text{avg}} \pi^2} c_0 = \bar{\sigma}_{\rv{\tilde{x}}}^2
		\label{eq:MSE_LB}
	\end{align}
	where $c_0 = 	- 3\gamma - 3 \log(2\pi) + 3 \Ci(2 \pi) - \pi^2 + 4 \pi \Si(\pi) - \pi \Si(2\pi)$, with $\gamma \approx 0.5772$ being the Euler-Mascheroni constant and $\Si(\cdot)$ and $\Ci(\cdot)$ being the sine- and cosine-integral functions, respectively.
	
	Furthermore, the autocorrelation function of the lowpass-distortion $\rv{\tilde{x}}(t)$ is given by	
	\begin{IEEEeqnarray}{rCl}
	s_{\rv{\tilde{x}\tilde{x}}}(\tau)  = \frac{1}{\pi} \int_{2 \pi W}^{\infty} S_{\rv{X}} (\omega) \cos(\omega \tau) d \omega
	\end{IEEEeqnarray}
	such that for its second derivative it can be written
	\begin{IEEEeqnarray}{rCl}
	s''_{\rv{\tilde{x}\tilde{x}}}(\tau) = \frac{\partial^2}{\partial \tau^2} s_{\rv{\tilde{x}\tilde{x}}}(\tau) = \frac{1}{\pi} \int_{2 \pi W}^{\infty} S_{\rv{X}} (\omega) \frac{\partial^2}{\partial \tau^2} \cos(\omega \tau) d \omega \label{eq:2nd_deriv_sxx(tau)}
	\end{IEEEeqnarray}
	where the exchangeability of differentiation and integration has been shown via Lebesgue's dominated convergence theorem \cite[Theorem 1.34]{Rudin1987} with the dominating function $g(\omega) = \omega^2 S_{\rv{X}}(\omega)$.
	For the derivation of the upper bound on $H(V_k)$ in (\ref{eq:H(Vk)}) we need an upper bound on $\mu$ and, thus, with (\ref{eq:mu}), (\ref{eq:var_z(t)}), and (\ref{eq:MSE(Sx(f))}) an upper bound on $S_\rv{X}(\omega)$. Due to $\frac{\partial^2}{\partial \tau^2} \cos(\omega \tau) \big|_{\tau=0} = - \omega^2$ in (\ref{eq:2nd_deriv_sxx(tau)}) and since $S_\rv{X}(\omega)$ is positive for all $\omega$, an upper bound on $S_\rv{X}(\omega)$ results in a lower bound on $s''_{\rv{\tilde{x}\tilde{x}}}(0)$ such that
	\begin{align}
	s''_{\rv{\tilde{x}\tilde{x}}}(0)  \ge - \frac{(1+2 c_1) \hat{P} }{T_{\text{avg}} \pi} \int_{{2 \pi W}}^{\infty} \omega^2 \left|G(\omega)\right|^2 {d \omega} = \underline{s}''_{\rv{\tilde{x}\tilde{x}}}(0)
	\end{align}
	yielding for the sine-waveform
	\begin{align}
	\underline{s}''_{\rv{\tilde{x}\tilde{x}}}(0) = - \frac{(1+2 c_1) \hat{P}}{2 T_{\text{avg}} \beta} \left[\pi^2 - \gamma - \log(2 \pi) - \pi \Si(2\pi) + \Ci(2 \pi)\right]= - \frac{(1+2 c_1) \hat{P}}{2 T_{\text{avg}} \beta} c_2 . \label{eq:S_xx_tilde_LB}
	\end{align}
% % % % % % % % % % % % % % % % % % % % % % % % % % % % % % % % % % % % % % % % % % % % % % % % % % % % % % % % % % % % % % % % % % % % % % % % % % % % % % % % % % % % % % % % % % % % % % % % % % % % % % % % % % % % %
% --------------------------------------------------------------------------------------------------------------------------------------------------------------------------------------------------------------------- %
% % % % % % % % % % % % % % % % % % % % % % % % % % % % % % % % % % % % % % % % % % % % % % % % % % % % % % % % % % % % % % % % % % % % % % % % % % % % % % % % % % % % % % % % % % % % % % % % % % % % % % % % % % % % %	

\section{Lower and Upper Bound on the Achievable Rate}
\label{sec:Conclusion}
\begin{figure}[t]
	\centering
	\includegraphics{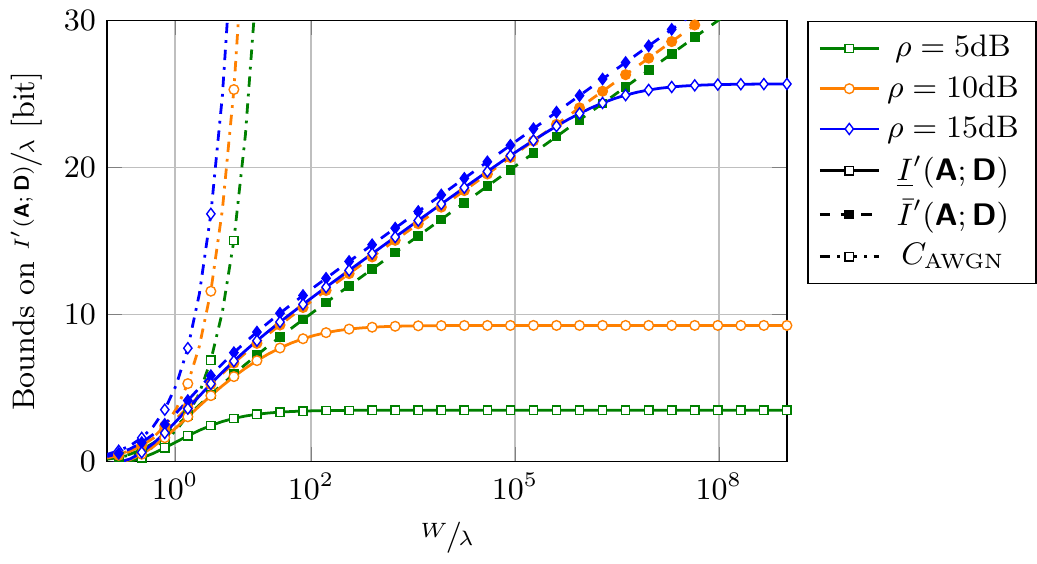}
	\caption{Lower and upper bound on $I'(\boldsymbol{\rv{A}};\boldsymbol{\rv{D}})$ in comparison to the AWGN capacity}
	\label{fig:results1}
\end{figure}
	Substituting (\ref{eq:T_avg}), (\ref{eq:I(A;D,V)=I(A;D_dach)}), (\ref{Rate_LB}), (\ref{eq:H(Vk)}), and (\ref{eq:entropy_rate_V}) into (\ref{eq:lowerBound_1}), a lower bound $\underline{I}'(\boldsymbol{\rv{A}};\boldsymbol{\rv{D}})$ on the mutual information rate of the 1-bit quantized time continuous channel is given by
	\begin{IEEEeqnarray}{rCl}
		\rule{-5mm}{0mm}\underline{I}'(\boldsymbol{\rv{A}};\boldsymbol{\rv{D}})
	\!=\!\frac{2 W \lambda}{2 W\!+\!\lambda} \left[\frac12 \log\left(\frac{e}{2 \pi}\right)\!+\!\frac12 \arcosh\!\left(\!\frac{2 \pi^2 W^2 \hat{P}}{\bar{\sigma}_{\rv{z}}^2 \lambda^2}\!+\!1\!\right)\!+\!\bar{\mu} \log\left(\!\frac{\bar{\mu}\!-\!1}{\bar{\mu}}\!\right)\!-\!\log(\bar{\mu}\!-\!1)\right] \label{eq:lb_I(AD)}
	\end{IEEEeqnarray}
	where $\bar{\sigma}_{\rv{z}}^2$, $\bar{\mu}$, $\underline{s}''_{\rv{zz}}(0)$ are obtained by applying $\bar{\sigma}_{\rv{\tilde{x}}}^2$ and $\underline{s}''_{\rv{\tilde{x}\tilde{x}}}(0)$, cf. (\ref{eq:MSE_LB}) and (\ref{eq:S_xx_tilde_LB}), to (\ref{eq:var_z(t)}), (\ref{eq:mu}), and (\ref{eq:def_s_zz(0)}).
	Furthermore, the corresponding upper bound for the given signaling scheme is\looseness-1
		\begin{IEEEeqnarray}{rCl}
				\bar{I}'(\boldsymbol{\rv{A}};\boldsymbol{\rv{D}}) 
				= \frac12 \int\limits_{-\frac12}^{\frac12} \log \left(1+\frac{(\nu-\underline{S}_\rv{\Delta}(f))^{+}}{\underline{S}_\rv{\Delta}(f)}\right) df \label{eq:ub_I(AD)}
		\end{IEEEeqnarray}
		with $\nu$, $\underline{S}_\rv{\Delta}(f)$, $\sigma_{\rv{A}}^2$, $\underline{\sigma}_{\rv{z}}^2$, and $\underline{\sigma}_{\rv{S}}^2$ are obtained by applying $\underline{\sigma}_\rv{\tilde{x}}^2$, cf. (\ref{eq:MSE_LB}), to (\ref{eq:cond_waterlevel}), (\ref{eq:S_Delta(f)}), (\ref{eq:input_variance}),  (\ref{eq:var_z(t)}), and (\ref{eq:sigmaS_Def}).
		
	 Fig.~\ref{fig:results1} shows the normalized upper and lower bound on the mutual information rate given in (\ref{eq:lb_I(AD)}) and (\ref{eq:ub_I(AD)}) as a function of $\sfrac{W}{\lambda}$ for different SNRs $\rho$. It results for constant SNR $\rho$ that $\sfrac{\underline{I}'(\boldsymbol{\rv{A}};\boldsymbol{\rv{D}})}{\lambda}$ becomes a function solely depending on the ratio $k=\sfrac{W}{\lambda}$\looseness-1
	  \begin{IEEEeqnarray}{rCl}
	  \frac{\underline{I}'(\boldsymbol{\rv{A}};\boldsymbol{\rv{D}})}{\lambda} = \frac{k}{2 k + 1} \left[\frac12 \log\left(\frac{e}{2 \pi}\right) + \frac12 \arcosh\left(2 \pi^2 k^2 f_1(k,\rho)+1\right) + f_2(k,\rho)\right] \label{eq:I(A;D)=f(rho,k)}
	\end{IEEEeqnarray}
	  where
	  \begin{align}
	  	f_1(k,\rho) = \frac{\hat{P}}{\bar{\sigma}_{\rv{z}}^2} = \frac{1+2 k}{\frac12 + 2 k} \frac{\rho}{1+\frac{(1+ 2 c_1(k)) c_0}{2 \pi^2 \left(\frac12 +2 k\right)} \rho}
	  \end{align}
	  with $c_1(k) = \frac{1}{\sqrt{1+4 \pi^2 k^2}-1}$. Furthermore,
	  \begin{IEEEeqnarray}{rCl}
	  	f_2(k,\rho) = (\bar{\mu}-1) \log(\bar{\mu}-1) -\bar{\mu} \log(\bar{\mu})
	  \end{IEEEeqnarray}
	  where
	  \begin{IEEEeqnarray}{rCl}
	  	\bar{\mu} = k \sqrt{\frac{\frac43 \pi^2 \left(\frac12 +2 k\right) +  2 (1+ 2 c_1(k)) c_2 \rho}{\pi^2 \left(\frac12 +2 k\right)+(1+ 2 c_1(k)) \frac{c_0}{2} \rho}} \exp \left(-\frac{f_1(\rho,k)}{2}\right)+1.
	  \end{IEEEeqnarray}
	 The same behaviour is exhibited by the upper bound on the mutual information rate. 
	
	Furthermore, it can be seen that the lower bound on the mutual information rate saturates for high bandwidths $W$ due to the limited randomness of the input signal controlled by $\lambda$. In the saturation range the average symbol duration $\rv{A}_k$ is large compared to the coherence time of the noise such that the expected number of additional zero-crossings within $\rv{A}_k$ becomes significant. In this case, the increase of the mutual information rate with side information $\underline{I}'(\boldsymbol{\rv{A}};\boldsymbol{\rv{D}},\boldsymbol{\rv{V}})$ with the bandwidth $W$ is compensated by the increase of $\bar{H}'(\boldsymbol{\rv{V}})$ representing the rate reduction due to additional zero-crossings. On the other hand, the upper bound on the mutual information rate $I'(\Vrv{A};\Vrv{D})$ grows without limits as there the additional zero-crossings are not considered. The saturation point of the lower bound depends on the SNR and until this points, both bounds are tight. 
	Moreover, in Fig. \ref{fig:results1} it can be observed that if $W$ is significantly smaller than $\lambda$, the lower bound becomes zero. Note that this does not mean that the mutual information rate is zero as (\ref{eq:lb_I(AD)}) is a lower bound.
	
\begin{figure}[h]
		\centering
		\includegraphics{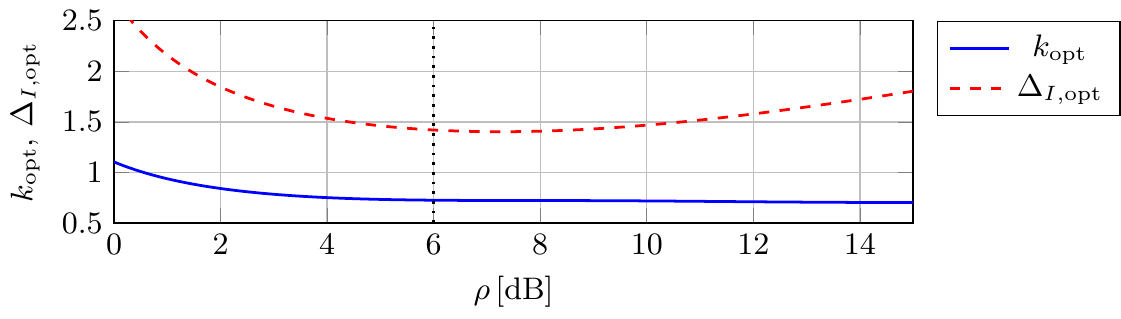}
		\caption{Optimal ratio $k=W \lambda^{-1}$ over the SNR and corresponding ratio $\sfrac{C_{\text{AWGN}}}{\underline{I}(\boldsymbol{\rv{A}};\boldsymbol{\rv{D}})}$, valid for the mid to high SNR regime $\rho \ge 6\,\text{dB}$}
		\label{fig:k_opt}
	\end{figure}

	For comparison the capacity $C_{\text{AWGN}} = W \log \left(1+ \rho \right)$ of the AWGN channel without output quantization is given, which represents an upper bound on the mutual information rate with 1-bit quantization. It can be seen that the lower bound is relatively tight for $\sfrac{W}{\lambda}$ in the order of 1.
	Hence, in order to avoid saturation of the mutual information rate for the chosen input distribution, the randomness of the input signal needs to be matched to the channel bandwidth, which is achieved by allowing $\lambda$ to grow linearly with $W$, i.e., $\lambda = \sfrac{W}{k}$ with $k$ being constant. 
	On a logarithmic scale $\log C_{\text{AWGN}}$ and $\log \underline{I}'(\boldsymbol{\rv{A}};\boldsymbol{\rv{D}})$ increase over $W$ with the same slope, leading to a constant offset given by
	\begin{align}
		\Delta_{I} &= \log C_{\text{AWGN}} - \log \underline{I}'(\boldsymbol{\rv{A}};\boldsymbol{\rv{D}}) = \log \left(\frac{C_{\text{AWGN}}}{\underline{I}'(\boldsymbol{\rv{A}};\boldsymbol{\rv{D}})}\right)\\
		& = \log \left[\frac{2 k+1}{2} \left(\frac{\log(1+\rho)}{\frac12 \log\left(\frac{e}{2 \pi}\right) + \frac12  \arcosh\left(2 \pi^2 k^2 f_1(k,\rho)+1\right) +f_2(k,\rho)}\right)\right] \label{eq:Delta}
	\end{align}
	which shows that there is a constant ratio between AWGN capacity and $\underline{I}'(\boldsymbol{\rv{A}};\boldsymbol{\rv{D}})$.
	The minimum of (\ref{eq:Delta}) w.r.t. $k$ is evaluated numerically and depicted in Fig. \ref{fig:k_opt}. In the high SNR regime the optimal $k$ is approximately $0.7$. 
	\begin{figure}[t]
		\centering
		\includegraphics{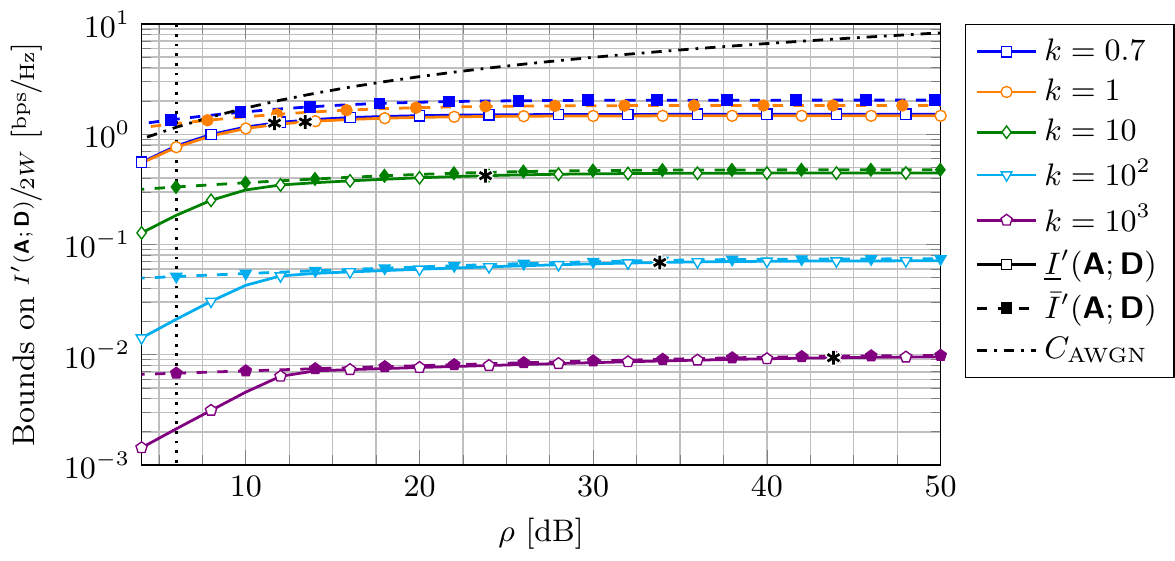}
		\caption{Lower and upper bound on $I'(\boldsymbol{\rv{A}};\boldsymbol{\rv{D}})$ normalized by the bandwidth $2 W$ and depending on the SNR $\rho$ and $k=\sfrac{W}{\lambda}$, valid for the mid to high SNR regime $\rho \ge 6\,\text{dB}$}
		\label{fig:results3}
	\end{figure}
	Note, that this optimum heavily depends on linking the transition time $\beta$ to the signal bandwidth $W$, cf. (\ref{eq:Relation_W_beta}). As can be seen in Fig.~\ref{fig:S_X(w)} in Section~\ref{subsec:Lowpass_A}, the utilization of the spectrum could be improved by reducing $W$, e.g., choosing $W=\sfrac{1}{T_{\text{avg}}}$. However, then the minimum symbol duration would not longer correspond to the coherence time of the noise, which is the basis of assumption~\ref{item:A2}. Thus, in order to improve the utilization of the spectrum the error event of deletions would have to be included in the model, see Appendix~\ref{app:deletions}.
	
	Furthermore, Fig. \ref{fig:results3} shows the lower bound ${\underline{I}}'(\boldsymbol{\rv{A}};\boldsymbol{\rv{D}})$ and upper bound ${\bar{I}}'(\boldsymbol{\rv{A}};\boldsymbol{\rv{D}})$ normalized by the bandwidth $2 W$ as a function of the SNR. This corresponds to the spectral efficiency of the given  signaling scheme. Due to the normalization by the signal bandwidth $2 W$, again the upper and the lower bound solely depend on the ratio $k=\sfrac{W}{\lambda}$, cf. (\ref{eq:I(A;D)=f(rho,k)}).
	We observe a gap between every lower and the respective upper bound on the mutual information rate that decreases with increasing SNR. Given that $\bar{I}'(\Vrv{A};\Vrv{D},\Vrv{V})$ and $\underline{I}'(\Vrv{A};\Vrv{D},\Vrv{V})$ do not diverge with increasing $\rho$, comparing (\ref{eq:upperBound_1}) and (\ref{eq:lowerBound_1}), it can be seen that in order for both bounds to become close, $\bar{H}(\Vrv{V})$ has to become smaller for increasing $\rho$ and, thus, the impact of symbol insertions has to be minor. As we have seen before that the impact of insertions decreases with the SNR, we conclude that the gap at low SNR results from symbol insertions that are not considered in the upper bound $\bar{I}'(\Vrv{A};\Vrv{D})$.
	The residual gap for large SNR remains roughly constant over the SNR and decreases with increasing $k$. One of the differences between  $\underline{I}'(\Vrv{A};\Vrv{D},\Vrv{V})$ and $\bar{I}'(\Vrv{A};\Vrv{D},\Vrv{V})$ is the bounding of the LP-distortion with $\bar{\sigma}_\rv{\tilde{x}}^2$ or $\underline{\sigma}_\rv{\tilde{x}}^2$, respectively. The LP-distortion is solely depending on $k$ and the signal-to-distortion ratio (SDR) is given as
		\begin{IEEEeqnarray}{rCl}
		2 \pi^2 \frac{(\frac12 + 2 k) (1+2 c_1(k))}{ c_0} = \frac{P}{\underline{\sigma}_\rv{\tilde{x}}^2} \le SDR \le \frac{P}{\bar{\sigma}_\rv{\tilde{x}}^2} = 2 \pi^2 \frac{\frac12 + 2 k}{(1+2 c_1(k)) c_0}.
		\end{IEEEeqnarray}
	In Fig. \ref{fig:results3}, black asterisks indicate for every $k$ the point, where $\rho = \sfrac{P}{\bar{\sigma}_\rv{\tilde{x}}^2}$. Although difficult to see, this is a good indicator where the saturation of $\underline{I}'(\Vrv{A};\Vrv{D})$ and $\bar{I}'(\Vrv{A};\Vrv{D})$ begins. From this point on increasing the SNR will have little to no benefit. The SDR decreases with increasing ratio $k=\sfrac{W}{\lambda}$ as then $\sigma_\rv{\tilde{x}}^2$ decreases, cf. Fig.~\ref{fig:S_X(w)}. Thus, the impact of the LP-distortion becomes small and upper and lower bound become tight.
	
	For the limiting case $\rho \rightarrow \infty$, the lower bound on the mutual information rate is
		\begin{IEEEeqnarray}{rCl}
			\lim\limits_{\rho \rightarrow \infty}\underline{I}'(\boldsymbol{\rv{A}};\boldsymbol{\rv{D}}) 
			&&= \frac{2 W}{2 k+1} \left[\frac12 \log\left(\frac{e}{2 \pi}\right) + \frac12 \arcosh \left( \frac{4 \pi^4 k^2(1+2k)}{(1+2 c_1(k)) c_0} + 1 \right)\right.\nonumber\\
			&&\rule{10mm}{0mm} \left. + \mu_g \log\left( \frac{\mu_g - 1}{\mu_g} \right) - \log(\mu_g - 1)\right]
		\end{IEEEeqnarray}
		with $\mu_g = \lim\limits_{\rho \rightarrow \infty} \mu = 2k \sqrt{\frac{c_2}{c_0}} \exp\left(-\frac{\pi^2 (1+2k)}{(1+2 c_1(k)) c_0}\right)+1$. It scales linearly with the bandwidth $W$.

\section{Conclusion}
\label{sec:conclusion}
We have derived a lower bound on the mutual information rate of the 1-bit quantized continuous-time AWGN channel focusing on the mid to high SNR regime. We furthermore have provided an upper bound on the mutual information rate of the specific signaling scheme used for deriving the lower bound. We have identified the parameter ranges in which both bounds are tight and have given explanations for those, in which they are not. 
As the lower and the upper bound, $\underline{I}'(\Vrv{A};\Vrv{D})$ and $\bar{I}'(\Vrv{A};\Vrv{D})$ are close in the mid to high SNR range, they provide a valuable characterization of the actual mutual information rate with the given signaling scheme on 1-bit quantized channels.
 
We have shown that in order to maximize the lower bound on the mutual information rate for a given bandwidth, the parameter $\lambda$ of the exponential distribution of the $\rv{A}_k$  controlling the randomness of the channel input signal needs to grow linearly with the channel bandwidth. For the given system model, the optimal coefficient $k = \sfrac{W}{\lambda}$ depends on the SNR and tends towards $0.7$ for high SNR. In contrast to the AWGN channel capacity, the bounds on the mutual information rate with 1-bit quantization saturate when increasing the SNR to infinity. This is due to the LP-distortion that is introduced because $\rv{x}(t)$ is not bandlimited. The high SNR limit of $\underline{I}'(\Vrv{A};\Vrv{D})$ and $\bar{I}'(\Vrv{A};\Vrv{D})$ scales linearly with the bandwidth for a given $k$ and the point of saturation in the SNR can be well approximated by the signal-to-distortion ratio of the signal.

\appendices
\section{Number of Zero-Crossings within a Transition Interval}
\label{app:ZC-count-Tk}
We want to verify the assumption \ref{item:A2} that within the interval $[T_k-\sfrac{\beta}{2},T_k+\sfrac{\beta}{2}]$ with very high probability only one zero-crossing occurs. This is a curve crossing problem depending on the deterministic waveform $f(t)$ and the random process $\rv{z}(t)$, which we approximated to be Gaussian with variance $\sigma_{\rv{z}}^2$ according to (\ref{eq:p_z(t)}). Hence, an equivalent way of looking at this problem is to study the zero-crossings of a non-stationary Gaussian process $\rv{q}(t) = \rv{z}(t) - \psi(t)$, where $\psi(t)$ is the deterministic curve to be crossed by $\rv{z}(t)$. For this purpose we define the transition interval $\mathbb{Y}=[0,\beta]$, where $y \in \mathbb{Y}$ is the time variable within the transition interval. Then the deterministic function
	\begin{IEEEeqnarray}{rCl}
	\psi(y) = -f\left(y-\frac{\beta}{2}\right)
	\end{IEEEeqnarray}
	depends on the waveform $f(t)$ of the transition. For the sine transition in (\ref{eq:f(t)_cosine}) it is given by
	\begin{IEEEeqnarray}{rCl}
		\psi(y) =  \cos\left(\frac{\pi}{\beta} y\right).
	\end{IEEEeqnarray}  
	The process $\rv{q}(t)$ has a zero-crossing only if $\rv{z}(y)=\psi(y)$. For the number of crossings $N_{T}(\psi)$ of a curve $\psi$ by a stationary Gaussian processes in the time interval of length $T$ it holds \cite{Kratz2006}
	\begin{IEEEeqnarray}{rCl}
\rule{-7mm}{0mm}	\E[N_{T}(\psi)]\!=\!\sqrt{-s''(0)}\!\int_{0}^{T}\!\!\varphi(\psi(y))\!\left[2 \varphi\!\left(\!\frac{\psi'(y)}{\sqrt{-s''(0)}}\!\right)\!+\!\frac{\psi'(y)}{\sqrt{-s''(0)}} \left(\!2 \Phi\left(\!\frac{\psi'(y)}{\sqrt{-s''(0)}}\!\right)\!-\!1\!\right)\!\right]\!dy \label{eq:Exp_ZC}
	\end{IEEEeqnarray}
	where $s(\tau)$ is the ACF of the Gaussian Process, $'$ denotes the derivative in time, i.e., w.r.t. $y$, and $\varphi$ and $\Phi$ are the zero-mean  Gaussian density and distribution functions with variance $\sigma^2_{\rv{z}}$, c.f. (\ref{eq:var_z(t)}), respectively. The variance of the number of zero-crossings is given by \cite{Kratz2006}
	\begin{IEEEeqnarray}{rCl}
	\Var(N_{T}(\psi))&&= \E[N_{T}(\psi)] - \E^2[N_{T}(\psi)] \nonumber\\
	&&\rule{5mm}{0mm}+ \int\limits_{0}^{T} \int\limits_{0}^{T} \int\limits_{\mathbb{R}} |\rv{q}'_{t_1}-\psi'_{t_1}||\rv{q}'_{t_2}-\psi'_{t_2}| \phi_{t_1,t_2}(\psi_{t_1},\rv{q}'_{t_1},\psi_{t_2},\rv{q}'_{t_2}) d \rv{q}'_{t_1} d \rv{q}'_{t_2} d t_1 d t_2\label{eq:Var_ZC}
	\end{IEEEeqnarray}
	where the subscripts $t_1$ and $t_2$ denote the time instants and $\phi$ is the multivariate zero-mean normal distribution of $\rv{q}(t_1)$, $\rv{q}'(t_1)$, $\rv{q}(t_2)$, and $\rv{q}'(t_2)$ with covariance matrix $\Sigma$
	\begin{figure}
		\centering
		\includegraphics{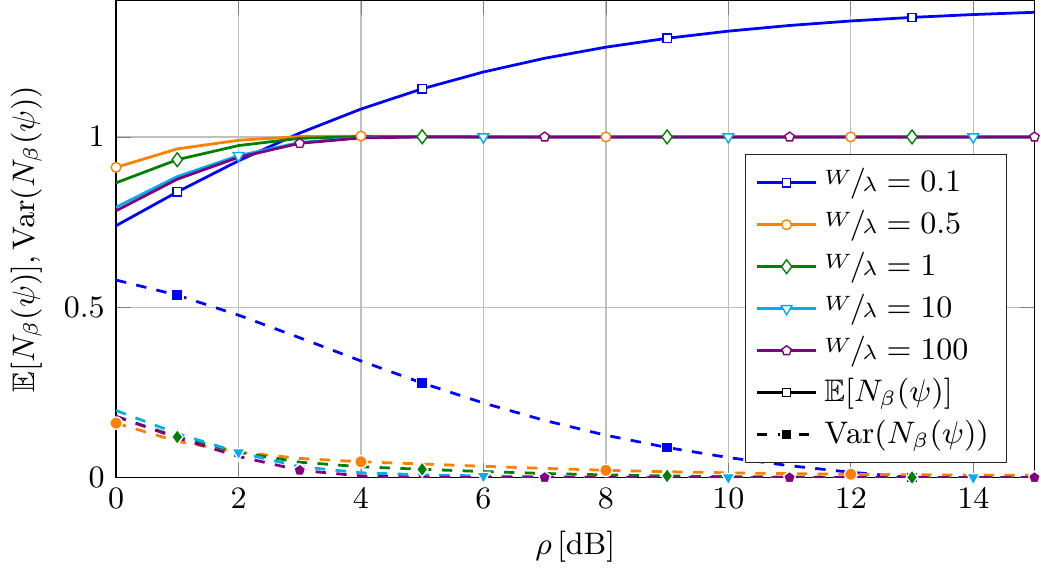}
		\caption{Expectation and variance of the number of zero-crossings in the transition interval $[T_k-\sfrac{\beta}{2},T_k+\sfrac{\beta}{2}]$, $\psi(y) = \cos\left(\frac{\pi}{\beta} y\right)$}
		\label{fig:E-Var(Nk)}
	\end{figure}
	\begin{IEEEeqnarray}{rCl}
	\Sigma = \begin{Bmatrix}
	s(0) & 0 & s(\tau) & s'(\tau) \\ 0 & -s''(0) & -s'(\tau) & -s''(\tau)\\ s(\tau) &  -s'(\tau) & s(0) & 0 \\ s'(\tau) & - s''(\tau) & 0 & -s''(0)
	\end{Bmatrix}.
	\end{IEEEeqnarray}
	The equations (\ref{eq:Exp_ZC}) and (\ref{eq:Var_ZC}) are evaluated and depicted in Fig. \ref{fig:E-Var(Nk)}. For $\sfrac{W}{\lambda} \ge 0.5$, the expectation of the number of zero-crossings converges to 1 for SNR $\ge$ 5\,dB while at the same time the variance converges to 0. Hence, for SNR $\ge$ 5\,dB almost surely exactly one zero-crossing exists in every transition interval. For $\sfrac{W}{\lambda} \ll 1$ the lower bound on the mutual information rate in (\ref{eq:lb_I(AD)}) becomes zero and, hence, the validity of the assumption is not relevant.

\section{Validity of the Gaussian Approximation}
\label{app:Gauss_Approx}
\begin{figure}
	\centering
	\includegraphics{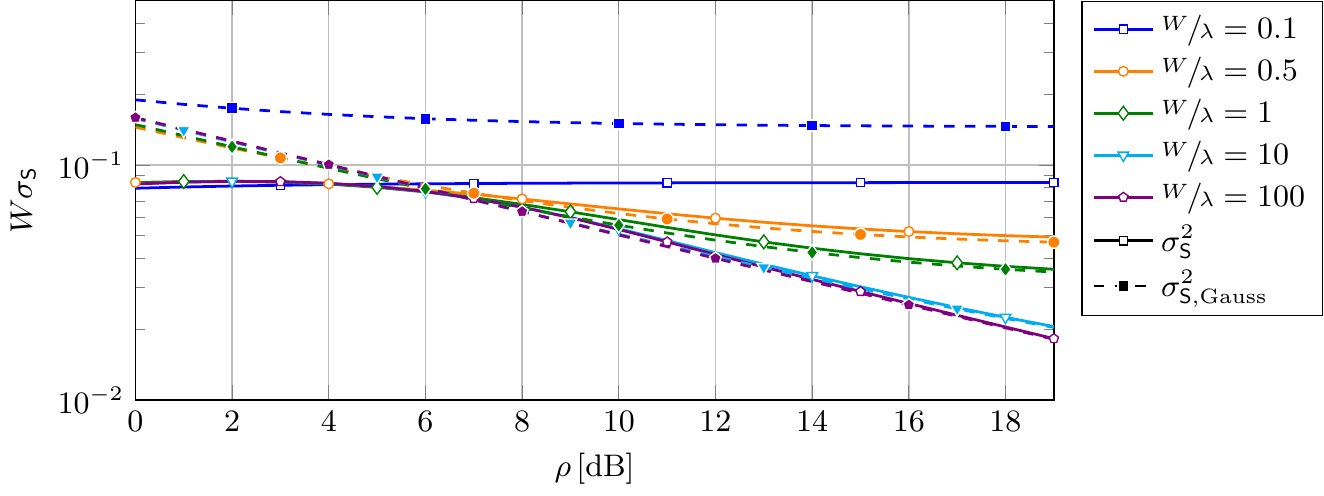}
	\caption{Normalized standard deviation of the original distribution $\sigma_{\rv{S}}$ (\ref{PDF_S}) and the Gaussian approximation $\sigma_{\rv{S},\text{Gauss}}$ (\ref{eq:PDF_S_Gauss})}
	\label{fig:Gauss_Approx}
\end{figure}

In order to quantify the high-SNR region for which the approximation of $p_\rv{S}(s)$ in (\ref{PDF_S}) by the Gaussian density in (\ref{eq:PDF_S_Gauss}) is valid, the variances of both densities have been evaluated and compared numerically. The corresponding standard deviations are depicted in Fig. \ref{fig:Gauss_Approx} and show a convergence of the variances in the relevant area of $\sfrac{W}{\lambda} \ge 0.5$ for SNRs larger 6\,dB. Comparing the variances is sufficient for our purpose as the further bounding of $I'(\boldsymbol{\rv{A}};\boldsymbol{\rv{D}},\boldsymbol{\rv{V}})$ is solely based on the variance of a Gaussian random process with equal covariance matrix.
% --------------------------------------------------------------------------------------------------------------------------------------------------------------------------------------------------------------------- %
\section{Upper-Bounding the Entropy of $\rv{V}_k$}
\label{app:deriv-H(Vk)}
The entropy maximizing distribution for a discrete, positive random variable with given mean $\mu$ is the geometric distribution \cite[Section 2.1]{kapur1989maximum}
$p_i = C q^i,~ i\in \mathbb{N}$
where $C = \frac{1}{\mu-1}$ and $q=\left(\frac{\mu-1}{\mu}\right)$. Thus, the entropy of $\rv{V}_k$ becomes
\begingroup
\allowdisplaybreaks
\begin{align}
H(V) & = - \sum_{i=1}^{\infty} p_i \log p_i = - C \log (C)\sum_{i=1}^{\infty} q^{i} - C \log (q) \sum_{i=1}^{\infty} q^{i} i \nonumber\\
& = - C \log (C) \frac{q}{1-q} - C \log (q) \frac{q}{(q-1)^2} \nonumber \\
& = (1-\mu) \log \left(\mu - 1\right) + \mu \log \mu.
\end{align} %
\endgroup
% --------------------------------------------------------------------------------------------------------------------------------------------------------------------------------------------------------------------- %

\section{Power Spectral Density of the Transmit Signal}
\label{app:spectrum_formula}
	When separating the spectrum $\rv{X}(\omega)$ of $\rv{x}(t)$ in (\ref{eq:S(x)}) into its real and imaginary part, we obtain
	\begin{IEEEeqnarray}{rCl}
	\rule{-2mm}{0mm}|\rv{X}(\omega)|^2 &&=\!\hat{P} \left(2 \pi \delta(\omega)\!+\!\Re \left\lbrace\! G(\omega)\!\sum\limits_{k=1}^{K} (-1)^k e^{-j\omega \rv{T}_k}\right\rbrace\!\right)^2\!+\!\hat{P} \Im^2 \left\lbrace \!G(\omega)\!\sum\limits_{k=1}^{K} (-1)^k e^{-j\omega \rv{T}_k}\right\rbrace \label{eq:E_XW_app}\\
	&&=\!4 \pi^2 \hat{P} \delta^2(\omega)\!+\!4 \pi \hat{P} \delta(\omega) \Re \left\lbrace\! G(\omega)\!\sum\limits_{k=1}^{K} (-1)^k e^{-j\omega \rv{T}_k}\!\right\rbrace\!+\!\hat{P}\left| G(\omega)\!\sum\limits_{k=1}^{K} (-1)^k e^{-j\omega \rv{T}_k}\right|^2\!\!\!.\label{eq:E_XW2_app}
	\end{IEEEeqnarray}
	In the second term of (\ref{eq:E_XW2_app}), all functions of $\omega$ can be shifted into the sum in order to obtain
	\begin{align}
	\delta(\omega) \Re \left\lbrace G(\omega) e^{-j\omega \rv{T}_k}\right\rbrace = \lim\limits_{\omega \rightarrow 0} \Re \left\lbrace G(\omega) e^{-j\omega \rv{T}_k}\right\rbrace=\beta. \label{eq:Exp_gamma_k_app}
	\end{align} 
	The third term of (\ref{eq:E_XW2_app}) can be written as
	\begin{IEEEeqnarray}{rCl}
		\hat{P}\left|G(\omega)\right|^2 \left|\sum\limits_{k=1}^{K} (-1)^k e^{-j\omega \rv{T}_k}\right|^2 = \hat{P} \left|G(\omega)\right|^2 \sum\limits_{k=1}^{K} \sum\limits_{v=1}^{K} (-1)^{k+v} \cos(\omega (\rv{T}_k-\rv{T}_v))
	\end{IEEEeqnarray}
	where
	\begin{IEEEeqnarray}{rCl}
	\left|G(\omega)\right|^2= \frac{2(1+\cos(\omega \beta))}{\omega^2}+a^2(\omega)+\frac{4 a(\omega)}{\omega} \cos\left(\frac{\omega \beta}{2}\right). \label{eq:F(w)_App}
	\end{IEEEeqnarray}
	Due to (\ref{eq:Exp_gamma_k_app}) being independent of $k$ and the alternating sign $(-1)^k$, the first two terms of (\ref{eq:E_XW2_app}) disappear in the limit in (\ref{eq:def_S(w)}). Exploiting the fact that the cosine is an even function, it remains for the PSD of the transmit signal in (\ref{eq:def_S(w)})
	\begin{align}
	S_\rv{X}(\omega) & = \frac{\hat{P} \left|G(\omega)\right|^2}{T_{\text{avg}}} \lim\limits_{K \rightarrow \infty} \frac{1}{K} \left(\sum_{k=1}^{K} (-1)^{2 k} + \sum_{k=1}^{K} \sum_{\substack{v=1 \\ v \ne k}}^{K} (-1)^{k+v} \E\left[\cos(\omega(\rv{T}_k-\rv{T}_v))\right]\right) \label{eq:inf_sum_app}\\
	&= \frac{\hat{P} \left|G(\omega)\right|^2}{T_{\text{avg}}} \left(1 + \lim\limits_{K\rightarrow \infty} 2 \sum_{n=1}^{K-1} (-1)^n \left(1-\frac{n}{K}\right) \E[\cos(\omega \rv{L}_n)]\right) \label{eq:S_X(w)_with-k_app}	
	\end{align}
	where $n=k-v$ is the index describing the distance between two arbitrary zero-crossing instances and $\rv{L}_n = \rv{T}_k-\rv{T}_v$ is the corresponding random variable with the probability distribution given in (\ref{eq:p(Xn)}),
	which results from the fact that $\rv{L}_n$ is the sum of $n$ consecutive input symbols
	\begin{IEEEeqnarray}{rCl}
		\rv{L}_n = \sum\limits_{i = 1}^{n} \rv{A}_{k+i}. \label{eq:L(A)}
	\end{IEEEeqnarray}
	As the input is i.i.d., it holds
	\begin{IEEEeqnarray}{rCl}
	p(\rv{A}_{k+1},...,\rv{A}_{k+n}) = p(\rv{A}_1,...,\rv{A}_{n}) = \prod\limits_{i=1}^{n} p(\rv{A}_i). \label{eq:p(A)_2}
	\end{IEEEeqnarray}
	From (\ref{eq:L(A)}) and (\ref{eq:p(A)_2}) we derive (\ref{eq:p(Xn)}).
% --------------------------------------------------------------------------------------------------------------------------------------------------------------------------------------------------------------------- %
\section{On the existence of a Coding Theorem}

\label{app:Coding_Theorem}
The existence of a coding theorem for memoryless insertion and deletion channels with finite discrete alphabet was proven by Dobrushin \cite{Dobrushin1967}. On the case of infinite and continuous alphabets and channels with memory, no results appear to be known yet.
The mathematical proof of a coding theorem is beyond the scope of this work, however, we will give an intuitive explanation for the achievability of the bound that we derived.

From a code perspective, in order to achieve an error probability converging to zero, it is required that no insertion pattern $\boldsymbol{\rv{D}}^{(\rv{V}_k)}$ can be interpreted as any subsequence of a valid codeword $[\rv{A}_1,\rv{A}_2,...,\rv{A}_k,...,\rv{A}_K]$. In our scenario, insertions are caused by excursions of the noise process $\rv{z}(t)$ beyond the signal levels $\pm \sqrt{\hat{P}}$. As both, the noise process and the signal $\rv{x}(t)$ are symmetric w.r.t. zero, it does not matter, if we consider the excursions $\rv{z}(t)>\sqrt{\hat{P}}$ or $\rv{z}(t)<-\sqrt{\hat{P}}$. 

Thus, the length of an inserted symbol is determined by the duration $\tau$ for which $\rv{z}(t)$ remains above $\sqrt{\hat{P}}$. To the best of our knowledge there are no easy to handle closed form results on the probability distributions of those excursion durations. However, the average length $\bar{\tau}$ of intervals the process spends above $\sqrt{\hat{P}}$ is given by \cite{Rice1957}
\begin{IEEEeqnarray}{rCl}
	\bar{\tau} = \pi \sqrt{\frac{\sigma_\rv{z}^2}{-s''_{\rv{zz}}(0)}} \exp \left(\frac{\hat{P}}{2 \sigma_\rv{z}^2}\right) \erfc\left(\sqrt{\frac{\hat{P}}{\sigma_\rv{z}^2}}\right) \label{eq:average_excursion_duration_app}
\end{IEEEeqnarray}
\begin{figure}
	\centering
	\includegraphics{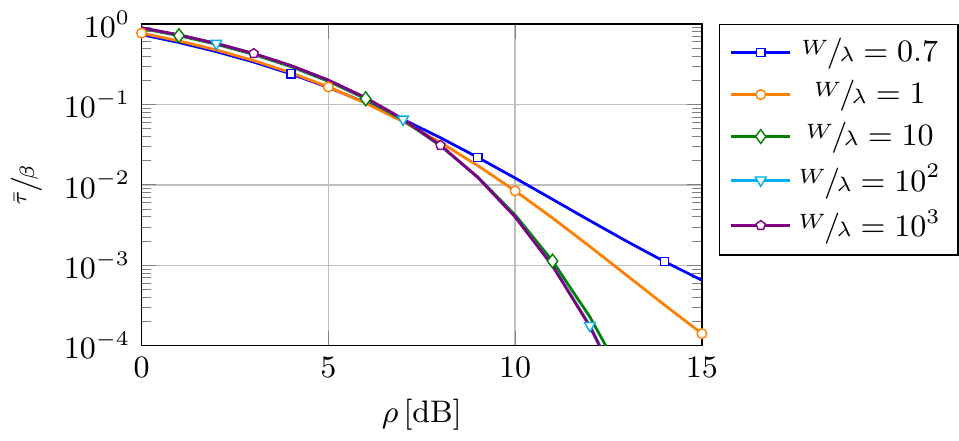}
	\caption{Average excursion duration normalized by minimum symbol length}
	\label{fig:Av_Excursion_Duration}
\end{figure}
Eq. (\ref{eq:average_excursion_duration_app}) is illustrated in Fig. \ref{fig:Av_Excursion_Duration}. It can be observed that the expected duration of insertions in the mid to high SNR domain, i.e., from approximately $6\,$dB, for which the bound we derived is applicable, is well below the minimum symbol duration $\beta$. Hence, the probability distributions of symbol patterns generated by insertions and original symbols differ significantly. This should allow for detection of a considerable amount of insertion, especially in the high SNR domain. If all insertions are detected, it remains a stationary colored Gaussian noise channel, for which a coding theorems exist. 
This gives an intuition that it is reasonable to assume the existence of a code such that our lower bound is achievable with diminishing error probability.

\section{Distribution of the Lowpass-Distortion Error}
\label{app:distr_LP-distortion}
At several places in the paper, we have assumed the additive distortion signal $\rv{z}(t)$ to be Gaussian distributed, namely
\begin{itemize}
	\item for derivation of the shifting error density $p_\rv{S}(s)$ in Section~\ref{subsec:genie-aided-rx_1}
	\item for the application of the Rice-formula and other level-crossing results in Section~\ref{sec:Aux_Proc}, Appendix~\ref{app:ZC-count-Tk}, and Appendix~\ref{app:Coding_Theorem}.
\end{itemize}
In order to justify this assumption, we have analyzed the  empirical probability distribution of $\rv{\tilde{x}}(t)$ by means of simulation: We generated $\rv{x}(t)$, applied the LP-filter and determined the histogram $h(\rv{\tilde{x}})$ with bin size $\delta=0.01 \max(\rv{\tilde{x}}(t))$. We evaluated the statistics of the signal $\rv{\tilde{x}}(t)$ over time with approximately $10^7$ time samples and the ensemble statistics at three points in time $t_1$, $t_2$, $t_3$ for $10^5$ samples each. We computed the empirical mean $\mu_{\rv{\tilde{x}},\text{em}}$ and the empirical variance $\sigma^2_{\rv{\tilde{x}},\text{em}}$ for each scenario.

\begin{figure}
	\includegraphics[width=\textwidth]{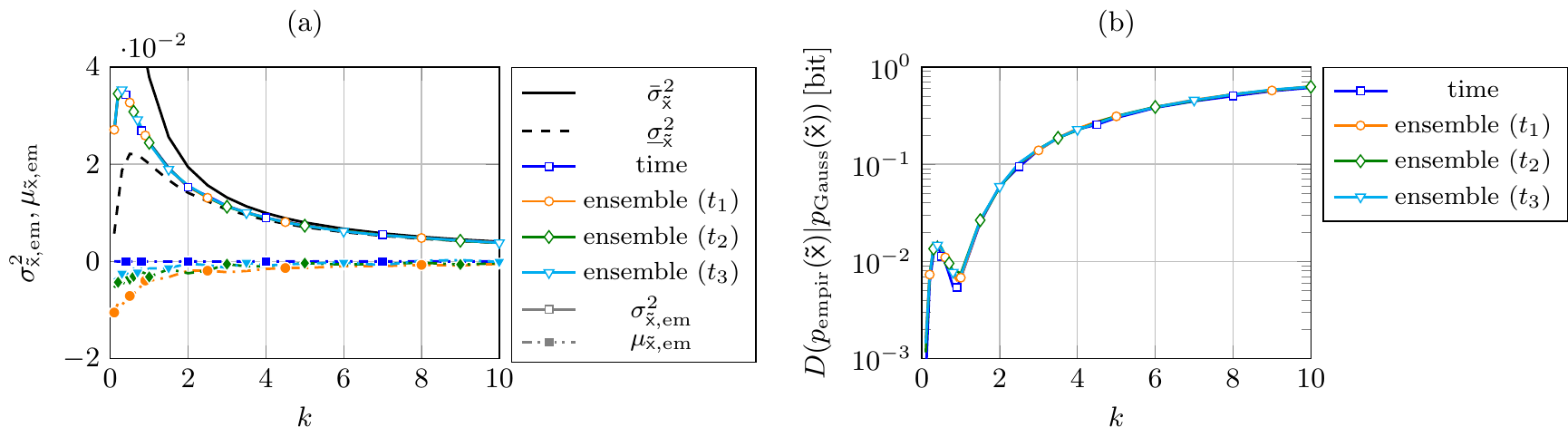}
	\caption{(a) Empiric mean $\mu_{\rv{\tilde{x}},\text{em}}$ and variance $\sigma^2_{\rv{\tilde{x}},\text{em}}$ of $\rv{\tilde{x}}$ and (b)  Kullback-Leibler divergence between empric probability distribution $p_{\text{empir}}(\rv{\tilde{x}})$ and Gaussian distribution $p_{\text{Gauss}}(\rv{\tilde{x}})$ with same mean and variance}
	\label{fig:Gaussian_distr_LP_distortion}
\end{figure}
In Fig.~\ref{fig:Gaussian_distr_LP_distortion} we see the results over $k=\sfrac{W}{\lambda}$. Part a) of the figure shows that the variances over time and over ensembles coincide. The mean values exhibit the same tendency, where time and ensemble averages converge with increasing $k$. Thus, we can approximate $\rv{\tilde{x}}(t)$ as wide sense stationary. Furthermore, the bounds on $\sigma^2_\rv{\tilde{x}}$ converge to the actual $\sigma^2_\rv{\tilde{x}}$ with increasing $k$.
In part b) of Fig.~\ref{fig:Gaussian_distr_LP_distortion} it can be seen that the Kullback-Leibler divergence $D(p_{\text{empir}}(\rv{\tilde{x}})||p_{\text{Gauss}}(\rv{\tilde{x}}))$ between the empirically obtained distribution $p_{\text{empir}}(\rv{\tilde{x}}) = h(\rv{\tilde{x}}) \delta^{-1}$ and the Gaussian distribution $p_{\text{Gauss}}(\rv{\tilde{x}}) \sim \mathcal{N}(\mu_{\rv{\tilde{x}},\text{em}},\sigma^2_{\rv{\tilde{x}},\text{em}})$ increases with $\sfrac{W}{\lambda}$. However, for $\sfrac{W}{\lambda}$ in the order of 1, which maximizes the lower bound on the mutual information rate for a given bandwidth $W$, $D(p_{\text{empir}}(\rv{\tilde{x}})||p_{\text{Gauss}}(\rv{\tilde{x}}))$ is relatively small and the Gaussian approximation is reasonable.

\section{Occurrence of Zero-Crossing Deletions}
\label{app:deletions}
Assumption \ref{item:A2} states the negligibility of deletions of zero-crossings based on the bandlimitation of the noise given in (\ref{eq:Relation_W_beta}). In this section, we verify this assumption by simulation. By removing the fixed relation of bandwidth $W$ and transition time $\beta$ in (\ref{eq:Relation_W_beta}) and allowing $W$ to take on any value, we augment the design space and potentially increase the spectral efficiency of the system, cf. Fig.~\ref{fig:S_X(w)}. However, the minimum distance between two zero-crossings $\beta$ is no longer linked to the coherence time of the noise either, which can potentially lead to deletion errors.

We consider a long sequence of $K=10^3$ symbols and a time resolution $\Delta t=10^{-3}$. For a given SNR, $\lambda$, and $\beta$, we generate $\rv{x}(t)$ and analyze the corresponding signal $\rv{r}(t)$ after the receive filter. Note that without (\ref{eq:Relation_W_beta}) the ratio $k=\sfrac{W}{\lambda}$ is not a property of the signal $\rv{x}(t)$ anymore as $W$ can take on any value. Thus, we define $\tilde{k}=\sfrac{1}{(2 \beta \lambda)}$ based on $\lambda$ and $\beta$ in such a way that $\tilde{k}=k$ if (\ref{eq:Relation_W_beta}) holds. In order to identify the locations of insertions and deletion, we match every received upcrossing\footnote{As upcrossing we denote zero-crossings with positive transition slope, i.e. from $-\sqrt{\hat{P}}$ to $\sqrt{\hat{P}}$. Correspondingly, downcrossings have a negative slope.} in $\rv{r}(t)$ to the closest upcrossing in $\rv{x}(t)$, likewise for the downcrossings, and count the deleted symbols.

\begin{figure}
	\includegraphics{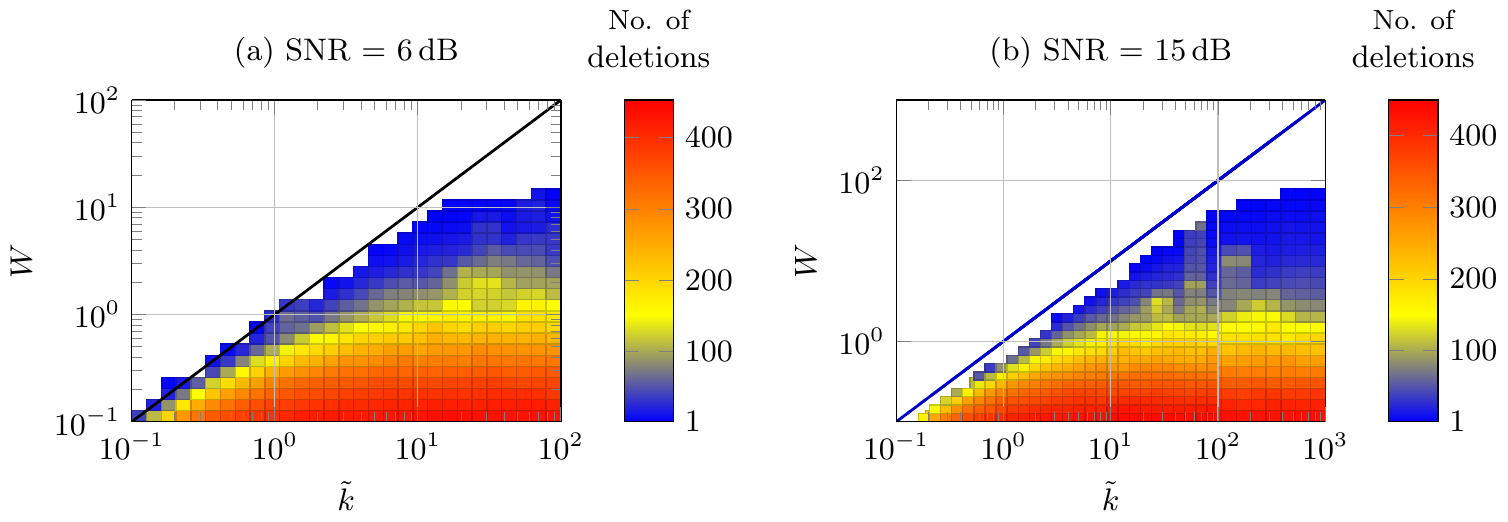}
	\caption{Number of deletions over bandwidth $W$ and over $\tilde{k}$, with $\lambda=1$ for (a) SNR=0\,dB and (b) SNR=15\,dB}
	\label{fig:no_of_insertions}
\end{figure}
The number of deletions is depicted in Fig.~\ref{fig:no_of_insertions} for two different SNR values of 6\,dB and 15\,dB. It can be seen that the SNR has a very small impact on the number of deletions occurred. The black line represents the case, when $W=\sfrac{1}{2 \beta}$. It can be seen that for bandwidths $W\ge\sfrac{1}{2 \beta}$, i.e., above the black line, no deletions occur as the dynamics of the noise are high compared to the minimum symbol duration $\beta$. However, when the bandwidth $W$ becomes smaller than $\sfrac{1}{2 \beta}$, deletions are possible and have to be considered in the system model - otherwise the spectral efficiency of the system will be overestimated.\looseness-1

% references section

\bibliographystyle{IEEEtran}

\bibliography{IEEEabrv,references}

\end{document}